\documentclass[fleqn,usenatbib]{mnras}

\usepackage{newtxtext,newtxmath}
\usepackage[T1]{fontenc}

\DeclareRobustCommand{\VAN}[3]{#2}
\let\VANthebibliography\thebibliography
\def\thebibliography{\DeclareRobustCommand{\VAN}[3]{##3}\VANthebibliography}

\usepackage{graphicx}	% Including figure files
\usepackage{appendix}
\usepackage{epstopdf}
\usepackage{natbib}
\usepackage{subcaption}
\usepackage{xspace}
\usepackage{color}
\newcommand{\kms}     {~km~s$^{-1}$\xspace}
\newcommand{\jy}      {~Jy~beam$^{-1}$\xspace}
\newcommand{\mjy}     {~mJy~beam$^{-1}$\xspace}
\newcommand{\mujy}    {~$\mu$Jy~beam$^{-1}$\xspace}

\newcommand{\msun}    {~$M_{\sun}$\xspace}

\newcommand{\cmt}     {~cm$^{-3}$\xspace}

\newcommand{\ceto}    {C$^{18}$O\xspace}
\newcommand{\tco}      {$^{13}$CO\xspace}

\newcommand{\dechms}[4]{$#1^{\rm h}#2^{\rm m}#3\mbox{$^{\rm s}\mskip-7.6mu.\,$}#4$} 
 
\newcommand{\decdms}[4]{$-#1^{\circ}#2'#3\mbox{$''\mskip-7.6mu.\,$}#4$}

\newcommand{\bo}[1]{{  #1}}
\newcommand{\com}[1]{{ #1}}
\newcommand{\refcom}[1]{{ #1}}

\title[Thackeray's globules revisited]{Thackeray's globules in IC~2944: the rocket effect revisited by ALMA}

\author[M. Fern\'andez-L\'opez \& L. A. Zapata et al.]{
M. Fern\'andez-L\'opez,$^{1,2}$
L. A. Zapata,$^{3}$\thanks{E-mail: l.zapata@irya.unam.mx}
B. Reipurth$^{4,5}$,
E. Santamaría$^{3}$,
M. Reiter$^{6}$,
P. Benaglia$^{1}$,
A. C. Raga$^{7}$\thanks{Deceased.}
\\
$^{1}$ Instituto Argentino de Radioastronom\'ia (CCT-La Plata, CONICET; UNLP; CICPBA), C.C. No. 5, 1894, Villa Elisa, Buenos Aires, Argentina \\
$^{2}$ Facultad de Ciencias Astron\'omicas y Geof\'isicas, Universidad Nacional de La Plata, Paseo del Bosque S/N, B1900FWA La Plata, Argentina\\
$^{3}$ Instituto de Radioastronomía y Astrofísica, Universidad Nacional Autónoma de México, C.P. 58089, Morelia, Michoacán, México\\
$^{4}$ Institute for Astronomy, University of Hawaii at Manoa, 640 N. Aohoku Place, HI 96720, USA\\
$^{5}$ Planetary Science Institute, 1700 E Fort Lowell Rd, Suite 106, Tucson, AZ 85719, USA\\ 
$^{6}$ Department of Physics and Astronomy, Rice University, 6100 Main St - MS 108, Houston, TX 77005, USA\\
$^{7}$ Universidad Nacional Aut\'onoma de M\'exico, Instituto de Ciencias Nucleares, A.P. 70-543, 04510, Ciudad de M\'exico, M\'exico
}

\date{Accepted XXX. Received YYY; in original form ZZZ}

\pubyear{\the\year{}}

\begin{document}
\label{firstpage}
\pagerange{\pageref{firstpage}--\pageref{lastpage}}
\maketitle

\begin{abstract}
The prominent Thackeray's globules are a collection of cloudlets seen in silhouette against the bright emission of the IC~2944 HII region, ionized by the Collinder~249 cluster of early-type stars \refcom{(placed at 2331$\pm$30~pc, derived from a Gaia DR3 analysis of the parallaxes of 11 massive stars)}. Here we present the analysis of Band~3 ALMA data that reveals the cold emission (continuum and molecular) associated with the neutral gas and its \refcom{kinematic} behavior. Many of the globules follow a linear velocity gradient that can be explained as the result of an acceleration process due to the 
%rocket effect originated by the photons ionizing material that shoots away from the surface of the globules. 
\refcom{rocket effect, where freshly ionized material streams away from the globule, compressing and accelerating it.} We identified 46 globules (12 of which are new detections), measured their kinematics, and estimated their densities and masses. At least 5 of them are associated with emission of dense molecular tracers and/or millimeter continuum sources and have indications of possible gravitational collapse. We 
%discuss \refcom{a} scenario where the globules are accelerated via \lq\lq rocket effect\rq\rq\,and 
\refcom{applied a simple model for the acceleration of the globules which accounts} for the observed kinematics. In this scenario
%, if resisting the ablation of small splinters and evaporation by the incident radiation, 
\refcom{only the most massive of the globules will be able to gravitationally collapse before being completely destroyed, in the process reaching} speeds of tens \kms, \refcom{and potentially} becoming low-mass walkaway/runaway protostars.

\end{abstract}

\begin{keywords}
ISM: clouds --  HII regions -- ISM: kinematics and dynamics --  photodissociation region (PDR) -- ISM: individual objects: Thackeray's globules
\end{keywords}

%%%%%%%%%%%%%%%%%%%%%%%%%%%%%%%%%%%%%%%%%%%%%%%%%%

%%%%%%%%%%%%%%%%% BODY OF PAPER %%%%%%%%%%%%%%%%%%

\section{Introduction}
\label{sec:intro}
The rocket effect, also called the %denominated 
\lq\lq interstellar rocket\rq\rq\, \citep{1954Spitzer,1955Oort,1990Bertoldi,1998Mellema,1998Johnstone,2005Raga,2009Henney,2020Reiter}, is a mechanism capable of accelerating interstellar clouds, cores, pockets, globules and/or shreds of dense gas \citep{1947Bok} by the action of stellar radiation from \refcom{early-type} stars. The radiation ionizes neutral material from the surface of a cloudlet and heats it up to thousands of degrees. This ionized and \refcom{heated} gas exerts an increased pressure and expands in every direction. 
%While the gas moving inwards the cloudlet is readily stopped, the gas \refcom{shooting} away from the cloudlet surface finds a low-density environment where it easily reaches the sound speed. 
\refcom{Gas moving toward the cloudlet is readily stopped, but gas expanding away from the cloudlet surface finds a low-density environment where it easily reaches the sound speed.} 
The gas moves predominantly toward the ionizing star and
%, following Newton's Third Law, 
the recoil accelerates the cloudlet in the opposite direction, in an analogous way that a %streaming-out 
jet accelerates a rocket. The more material streaming off the cloudlet's surface, the faster the cloudlet/rocket becomes. 
%So much so that if they survive destruction due to photoionization, some cloudlets could reach velocities of several tens of \kms \citep{1955Oort}.   
Cloudlets that survive destruction due to photoionization could reach velocities of several tens of \kms \citep{1955Oort}.   

Approximately 20--30\% of O-type stars and 5--10\% of B and Be stars are runaways with high peculiar velocities \citep[][ and references therein]{2011Tetzlaff,2024Stoop,2023CarreteroCastrillo}. A major avenue to create these runaways is through dynamical encounters \citep{1967Poveda} and less so through a sudden loss of mass of one of the members of a binary system, which leaves  the other star gravitationally unbound, as may occur when a supernova explodes in a binary \citep{1961Blaauw}. While high-mass runaways are easier to detect observationally, numerical simulations show that low-mass runaways should be the dominant fraction of the runaway population \citep[e.g., ][]{2019Farias,2025Fajrin}. The alternative pathway of creating runaway stars through interstellar rocket acceleration is usually discarded for the most massive and extreme-velocity runaways because it \lq\lq would require improbably large collections of O-type stars\rq\rq\, \citep[in the words of ][]{1961Blaauw}.
For lower mass stars reaching not extremely high velocities, this may be a reasonable avenue since the required initial mass of the original cores may not be too high \citep[e.g.,][]{1955Oort}. However, this idea has gained only modest support \citep[\refcom{but see}][and references therein]{2015Haworth}.  

Thackeray's globules are a collection of small dark clouds first reported by \cite{1950Thackeray}. They comprise an ensemble of globules and splinters that are seen in silhouette, appearing dark \refcom{at} optical wavelengths against a bright background \citep{1997Reipurth,2003Reipurth} produced by the emission \refcom{from the early-type stars} of the Collinder~249 cluster. 
These stars are also responsible for the IC~2948/IC-2944 HII region, a luminous stellar complex at the inner edge of the Carina spiral arm \citep{2011Sana}. \cite{2003Reipurth} argued that the Thackeray globule complex, located in the foreground of the illuminating O-type stars, are the remnants of an elephant trunk that underwent strong photo-erosion and ablation, that fragmented it into pieces now \refcom{on} their way to destruction. No infrared or protostellar sources have been found within the globules so far. Finally, the overall kinematics of the globules, observed using single-dish CO observations, presents a wide range of velocities (spanning $\sim20$\kms), which led \cite{1997Reipurth} to suggest Rayleigh-Taylor instabilities in a dense shell pushed away by the expanding HII region as its possible cause.

\subsection{Gaia DR3 distance to Collinder~249}
\label{sec:distance}
%\appendix

%\section{Gaia DR3 distance analysis} \label{appendix}
%This Appendix shows a table with parallaxes compiled from searching the Gaia DR3 archive (Table \ref{t:distances}), and a Figure showing the position of the OB stars used to derive the average distance of 2331$\pm30$~pc to Collinder~249 (Figure \ref{f:distances}).

\begin{figure*}
\centering
\includegraphics[width=1.\linewidth]{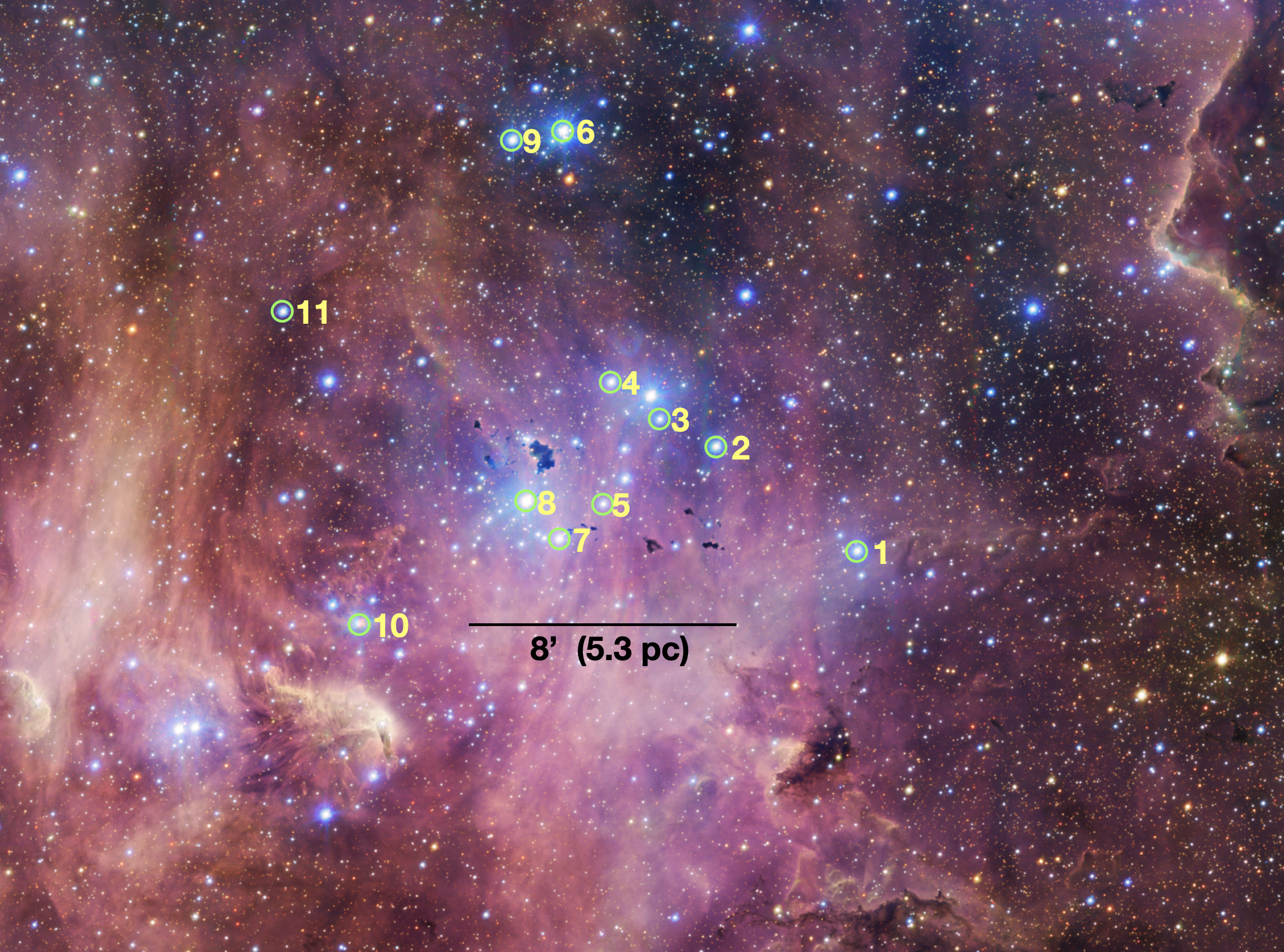}
\caption{Optical composite image from the VLT Survey Telescope (VST\refcom{, where VLT stands for the Very Large Telescope at Cerro Paranal, in Chile}) showing the Thackeray's globules. The numbers associated with the early-type stars marked in the image correspond to the identification numbers in Table \ref{t:distances}. HD~101205 and HD~101191, the dominant sources of photoionizing radiation, are numbered 8 and 5, respectively. \com{The scale-bar indicates the approximate linear extent of the group of globules assuming a distance of 2331~pc. Credit: ESO/VPHAS+ team. Acknowledgement: CASU. https://www.eso.org/public/images/eso2320a/}}
\label{f:distances}
\end{figure*}

IC~2944 and IC~2948 are part of a single large HII region in Centaurus, which is traditionally labeled IC~2944, although in fact the main region is IC~2948, while IC~2944 is associated with a major cloud and its bright rim \citep[see][]{2008Reipurth}. We will continue the standard practice of using IC~2944 for the entire complex.
There has been a lot of debate about the distance to the Collinder~249 cluster, centered on the star HD~101205, the brightest in the surroundings of the Thackeray globule complex along with HD101191 \citep[e.g.,][]{2011Sana,2017Krelowski}. \cite{1980Ardeberg} carried out a detailed study of OB stars seen towards IC~2944, and based on spectrographic and photometric data they distinguished several groups at different distances. They found that the rich and compact aggregate of young stars is located at a kinematic distance of 1750~pc. \cite{2014Baume} favored a distance of roughly 2300~pc, while \cite{2017Krelowski}, based on observations of the [CaII] interstellar line, determined a distance of 1600~pc.
We have selected OB-stars within $\sim$10~arcmin of the largest globule and have used Gaia DR3 \citep{2023Vallenari} to determine their distances (Figure~\ref{f:distances}). There is indeed a range of distances from 1.6 to 2.6 kpc, with a subgroup of 11 stars that cluster around a weighted mean distance of 2331$\pm$30~pc (Table~\ref{t:distances}), which we adopt here as the distance to the globules. At this distance one arcsecond corresponds to about 0.01 pc, so the globule complex has a projected extent of about 5~pc, and the largest globule has the approximate dimensions of 0.4$\times$0.7~pc$^2$.

\begin{table}
\caption{Massive stars associated with Thackeray's globules}
\label{t:distances}
\begin{tabular}{@{}lllll@{}}
\hline\hline
\# & Star & SpT & Parallax & Distance \\
   &      &     & [mas]    & [pc] \\   
\hline \hline
1 & HD 101008    & O9V    & 0.41/0.02 &  2439/119    \\
2 & HD 101084    & B1V    & 0.46/0.02 &  2174/95     \\
3 & CPD -62 2153 & B1IV   & 0.41/0.02 &  2439/119    \\
4 & HD 308813    & O9.7IV & 0.40/0.02 &  2500/126    \\
5 & HD 308819    & A0     & 0.45/0.02 &  2222/99     \\
6 & HD 101190    & O6IV   & 0.42/0.02 &  2381/114    \\                  
7 & HD 101191    & O8V    & 0.45/0.02 &  2222/99     \\
8 & HD 101205    & O7III  &  - - -    &  - - -$^{a}$ \\
9 & HD 101223    & O8V    & 0.42/0.02 &  2381/114    \\
10 & HD 101298    & O6V    & 0.43/0.02 &  2326/109    \\
11 & HD 308809    & B8     & 0.43/0.01 &  2326/54     \\
\hline
\multicolumn{5}{p{0.75\linewidth}}{\raggedright $^a$ HD 101205 is saturated in the Gaia data}\\
\end{tabular}
\end{table}

In this work we present new Band 3 Atacama Large Millimeter/submillimeter Array (ALMA) and archival Hubble Space Telescope (HST) observations covering the northeastern part of the complex harboring the Thackeray's globules. In Section \ref{sec:observations} we describe the observations. Section \ref{sec:results} shows the main results from the 2.9~mm continuum and the molecular line emission. We analyze and discuss the kinematics found in the individual globules and splinters, and explore the hypothesis that they have been accelerated by rocket effect in Section \ref{sec:discussion}. Finally, Section \ref{sec:summary} briefly summarizes our main findings.

%There has been a lot of debate about the distance to the Collinder~249 cluster, centered on the star HD~101205, the brightest in the surroundings of the Thackeray globule complex along with HD101191 \citep[e.g.,][]{2011Sana,2017Krelowski}. Throughout this work, we adopt 1800~pc as the distance to Collinder~249 \citep{1980Ardeberg, 2005Kharchenko}, in accordance with the value used by \cite{1997Reipurth}, as an intermediate value between extremes that range between 1600~pc and 2300~pc in the literature. We note that the recent estimate of 1560~pc by \cite{2017Krelowski} based on observations of the [CaII] interstellar line could be a more robust estimate, though.

\section{Observations and imaging}
\label{sec:observations}

\subsection{ALMA Band 3 observations}
\label{sec:almaobs}

 ALMA observations of Thackeray's globules were carried out on 11 January 2016 to 16 January 2025 as part of Cycle~3 program 2015.1.00908.S (PI: Bo Reipurth). The observations were performed in one session using from 43 to 46 antennas with diameters of 12\,m, resulting in projected baseline lengths ranging from 24.3 to 133.6\,m (equivalent to 8.1--44.5\,k$\lambda$). Given the $63\farcs8$ full width at half maximum (FWHM) of the Band 3 primary beam, the millimeter continuum and spectral line emission associated with the Thackeray's globules could not be fully captured in a single pointing. Therefore, a small mosaic of 19 Nyquist-sampled pointings was used to ensure the spatial coverage of those globules mapped in previous Hubble Space Telescope observations (see Section \ref{sec:hstobs}).
A total on-source integration time of 24 minutes was achieved across the observing session.
The maximum recoverable scale for this observation is $24\farcs3$. After retrieving the data from the archive, a standard calibration provided by the observatory was applied using CASA. 

While in this publication we focus on the continuum emission (at 104.6~GHz, or 2.9~mm) and the \tco~(1-0) emission (rest frequency, 110.20135~GHz), there is a set of lines with positive detections among the ensemble of Thackeray's globules. The lines, that appear in other spectral windows of the ALMA frequency setup, are reported in Table~\ref{t:lineid} along with their rest frequencies.
We prepared a continuum image with natural weighting that has a synthesized beam of $3\farcs0\times2\farcs9$ and a position angle of $-6\fdg0$. \bo{The rms noise level of this image is $21$\mujy. We also prepared spectral line images with a robust parameter of 0.5 and beam sizes ranging between $2\farcs4$ and $2\farcs7$  depending on the spectral window. For spectral windows 2 and 3 (the latter contains the \tco line), the typical rms noise level in a 0.08\kms channel width is 5.6\mjy, whereas for spectral windows 4, 5, and 6 (which have coarser spectral resolution) the rms noise level in a 0.19\kms channel is 2.8\mjy.}  

\subsection{Archival HST observations}
\label{sec:hstobs}
Optical observations were retrieved from the Hubble Space Telescope (HST) archive. The data were taken using the H$\alpha$ filter of the planetary camera chip of the WFPC2 at 656~nm on February 7, 1999. The data were presented in \cite{2003Reipurth} and we refer to this work for further details.

%\subsection{Infrared data??}
%\label{sec:irobs}

\section{Results}
\label{sec:results}

\subsection{Globule ID and overall kinematics}
\label{sec:globuleid}
\begin{figure}
\centering
\includegraphics[width=1.15\linewidth]{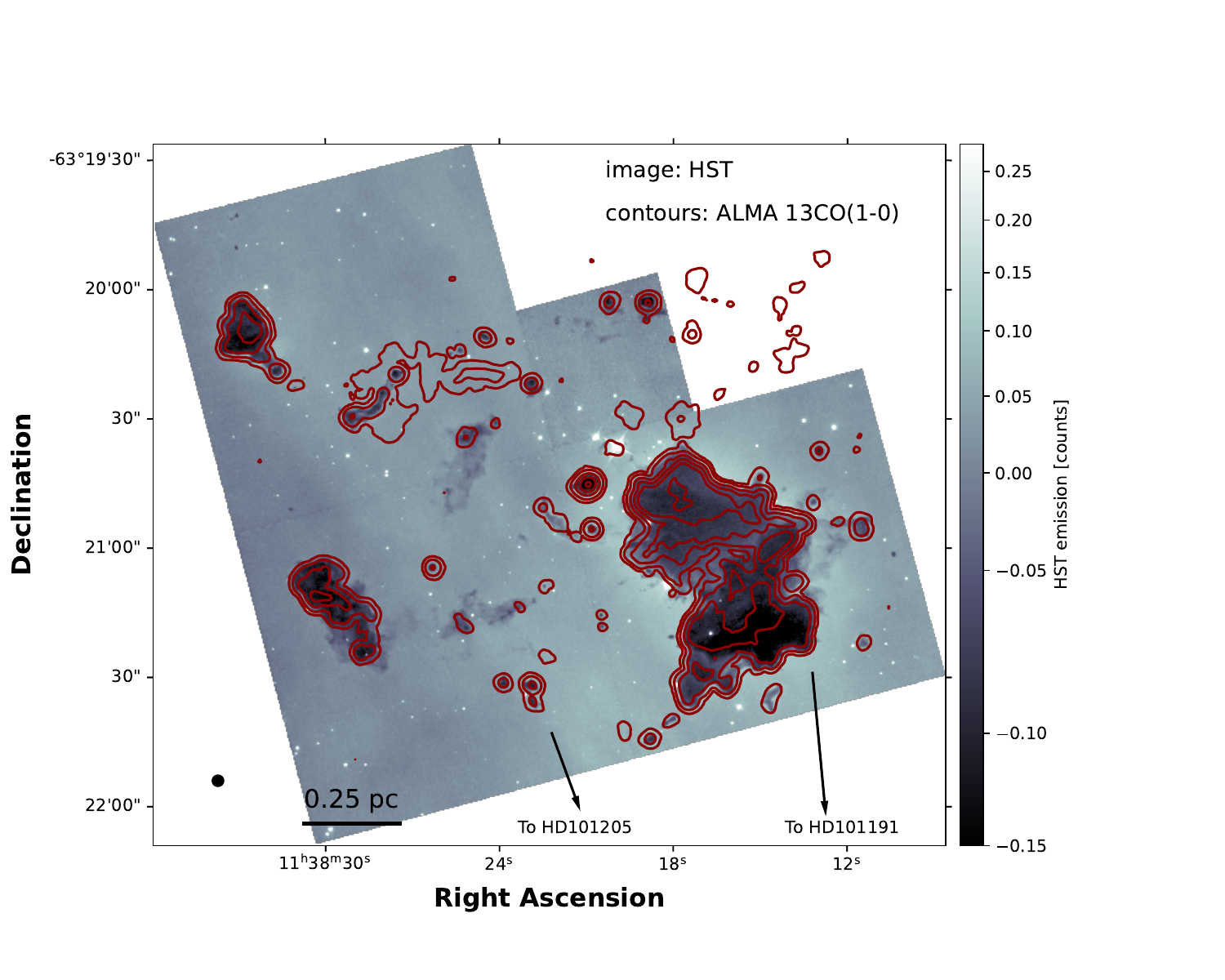}
\caption{HST image in arbitrary count units overlaid with \tco peak emission contours (moment~8) at levels 0.03, 0.1, 0.3, 0.6, 0.9, and 1.2\jy. \bo{Note that the moment~8 image is built finding the peak intensity of the spectrum associated with each spatial pixel of a velocity cube.}}
\label{f:13co}
\end{figure}

A detailed inspection of the spatial distribution of the \tco(1-0) emission (Figure \ref{f:13co}) revealed a collection of sources seen at different line-of-sight \refcom{LSR} velocities in the observed field\footnote{\refcom{Throughout this paper we define $v_{\rm lsr}$ as the observed line-of-sight velocity, $v_{\rm cloud}$ as the average velocity of the remnants of the original molecular cloud, and $v_{\rm rel}=v_{\rm lsr}-v_{\rm cloud}$ as the relative velocity to the cloud kinematic system.}}. Comparing these sources with the catalog of globules seen in silhouette and reported in \cite{1997Reipurth} and \cite{2003Reipurth}, 
%we uncovered 13 new globules and spectrally separated most of the already known. 
\refcom{we recovered most of the already known and discovered 13 new ones.}
Table~\ref{t:globules1} summarizes our findings with \refcom{the 
%average position 
coordinates of the barycenter of the polygons surrounding every identified source,} the Gaussian fit results of their \tco spectra \refcom{(line center LSR velocities)}, their integrated brightness temperatures, column densities, and masses. The entries of the table are arranged in three kinematic blocks of sources separated by their line-of-sight velocities (B1, B2, and B3, see below) and then by right ascension within each block. In the first block (B1), the line-of-sight velocities of the globules range between $-$20\kms and $-$34\kms; the second block (B2) contains only three globules with velocities between $-$7\kms and $-$11\kms; the third block (B3) comprises the most red-shifted globules with line-of-sight velocities between 45\kms and 49\kms. 

%Some globules 
\refcom{About 10\% of the globules} detected \refcom{at} optical wavelengths are undetected here up to the sensitivity threshold of the ALMA observations. Of these, Th19, Th21, and Th42 lie close to the less sensitive edges of the mosaic; Th39, Th41, and some diffuse portions of Th12, are probably too weak to be detected. The new detected globules (labeled as ALMA sources in Table \ref{t:globules1}) are mainly detected on the more red-shifted blocks B2 and B3. Among the globules seen \refcom{at} optical wavelengths, only Th40 is in a red-shifted block. The optical detection may be associated with a blue-shifted structure undetected by ALMA.

\begin{figure}
\centering
\includegraphics[width=\linewidth]{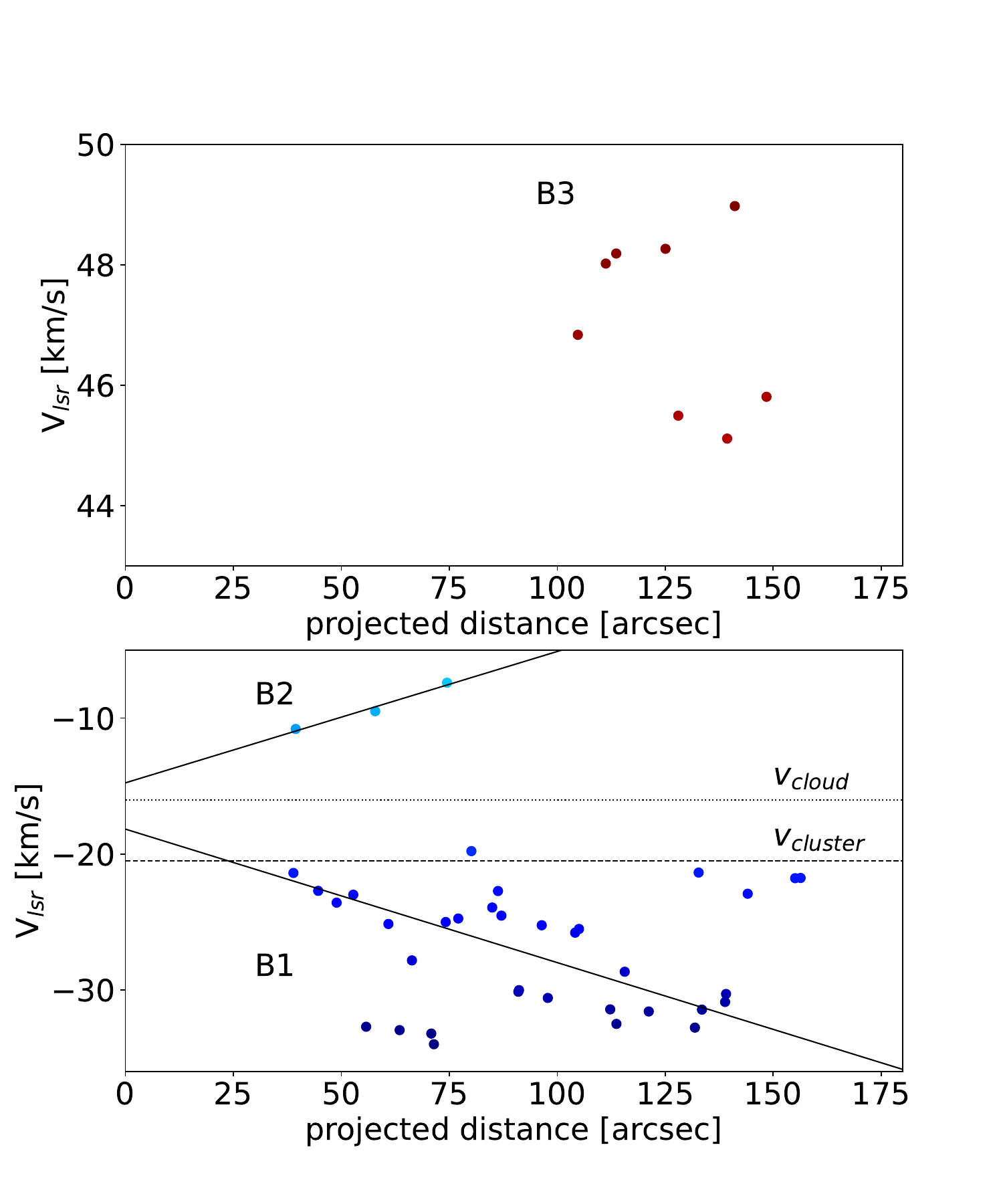}
\caption{Position-velocity plots of the Thackeray's globules. Block B3 (upper pannel) and blocks B1 (most blue-shifted globules) and B2 (intermediate velocity globules; lower pannel) are labeled accordingly. Projected distance is measured from the star HD~101205. Colors represent the line-of-sight velocity, $v_{\rm lsr}$. Black \refcom{solid} lines are two separate fits to part of the globules of block B1 (except Th02 and Th03) and the three globules of block B2. \refcom{Two horizontal dashed lines mark the LSR velocities of the stellar cluster and the cloud emission from gas remnants of the original molecular cloud (see Section \ref{sec:globalkinematics}).}}
\label{f:pv}
\end{figure}

Figure \ref{f:pv} shows two position-velocity diagrams for globules in blocks B3 (upper pannel) and B1 and B2 (lower pannel), centered on the position of HD~101205, the dominant ionizing star in the surroundings. We found that globules in blocks B1 and B2 seem to roughly follow linear velocity gradients (note that to fit the data of block B1, we did not take into account globules Th02 and Th03). These velocity gradients \refcom{almost overlap} at velocities -15\kms and -18\kms at the position of the reference star. The slope of the velocity gradient is the same for both blocks, and corresponds to a timescale of about 830~yr. In Section \ref{sec:globalkinematics} we discuss these velocity gradients as possibly due to a rocket effect acceleration process of the globules.

\subsection{Physical characterization of globules}
\label{sec:globuleid}
The density and mass of the globules were estimated from the \tco emission by making the typical assumptions that emission is optically thin and in local thermodynamic equilibrium (LTE; see Table \ref{t:globules1}). We adopted \bo{a range of excitation temperatures between 11~K and 19~K, based upon the brightest optically thick spots in the \tco~(1-0) emission, in agreement with the value derived in \cite{1997Reipurth}.} We used equation 80 in \cite{2015Mangum}, adapted for the \tco~(1-0) line, to derive the \tco column density. We also used a \tco abundance of $1.4\times10^{-6}$ (\citealt{2021Benedettini} based on the classical study of \citealt{1982Frerking}) to estimate molecular densities and a factor 2.8 as the mean molecular weight \citep{2012Crutcher} to derive the masses. \bo{Assuming spherical globules, their median diameter is 0.048~pc. Their median density and mass is 2.8-3.6$\times10^3$\cmt and 0.02-0.03\msun, respectively, and the \refcom{total mass} of all the globules is 34.6-44.4\msun. \refcom{88\% of the} mass content is harbored in Thackeray-1 (here labeled Th01), 5\% in Th02, 4\% in Th03, and 1\% in Th14. However, we make the caveat that \tco emission is moderately optically thick ($\tau_{\rm 13CO}\approx 1$) in a few spots of Th01A, Th01B, Th01C, Th02A, and Th14, which may imply density and mass underestimations of factors 1.1-2.3 in these few lines-of-sight.}

\begin{table*}
\caption[]{Gaussian fits results of globules.}
\label{t:globules1}
%\centering
\small
\begin{tabular}{@{}lcccccccc@{}}
\hline\hline
Label &	RA	& DEC	& Amplitude	&	Center	& FWHM	& Integral & \bo{$N_{^{13}\rm CO}$} & \bo{Mass}	\\
      & [hms] & [dms] & [K] & [\kms] & [\kms] & [K\kms] & [$10^{14}$~cm$^{-2}$] & [\msun]\\
\hline \hline
\multicolumn{9}{c}{Block B1} \\
\hline
%Th02	& 11:38:32.654 & -63:20:10.68 & 173$\pm$1 & -21.775$\pm$0.004 & 1.324$\pm$0.009 & 488$\pm$3 \\Th02A	& 11:38:32.803 & -63:20:09.82 & 201$\pm$1 & -21.758$\pm$0.004 &	1.28$\pm$0.01 & 547$\pm$4 \\
Th02A$^a$	& 11:38:32.803 & $-$63:20:09.82 & 3.52$\pm$0.02 & $-$21.758$\pm$0.004 & 1.28$\pm$0.01 & 4.78$\pm$0.06 & 43.6-56.0 & 1.81-2.33 \\
Th02B$^a$	& 11:38:31.762 & $-$63:20:18.43 & 1.12$\pm$0.02 & $-$22.918$\pm$0.006 & 0.85$\pm$0.02 & 0.51$\pm$0.02 & 4.6-5.9 & 0.04-0.05 \\
%Th03	& 11:38:29.442 & -63:21:14.43 & 88.5$\pm$0.8 & -30.126$\pm$0.006 & 1.33$\pm$0.01 & 251$\pm$2 \\
Th03B$^a$	& 11:38:29.682 & $-$63:21:06.79 & 1.65$\pm$0.01 & $-$30.584$\pm$0.006 & 1.61$\pm$0.01 & 2.83$\pm$0.05 & 25.8-33.1 & 0.49-0.64  \\
Th03A$^a$	& 11:38:29.603 & $-$63:21:15.21 & 1.63$\pm$0.02 & $-$30.011$\pm$0.006 & 1.15$\pm$0.01 & 1.99$\pm$0.04 & 18.1-23.3 & 0.951.23 \\
Th11$^b$	& 11:38:28.331 & $-$63:20:25.19 & 0.47$\pm$0.01 & $-$29.33$\pm$0.01 & 0.89$\pm$0.02 &   &  &   \\
    &     &       & 0.42$\pm$0.01 & $-$28.27$\pm$0.01 & 0.83$\pm$0.02 & 0.42$\pm$0.01 & 3.8-4.9 & 0.13-0.17  \\
Th33	& 11:38:26.297 & $-$63:21:04.70 & 1.27$\pm$0.02 & -24.546$\pm$0.004 & 0.509$\pm$0.009 & 0.69$\pm$0.02 & 6.3-8.1 & 0.04-0.06  \\
Th38	& 11:38:25.312 & $-$63:20:14.29 & 0.92$\pm$0.05 & -32.77$\pm$0.02 & 0.57$\pm$0.04 & 0.56$\pm$0.06 & 5.1-6.5 & 0.01  \\
Th34A$^c$	& 11:38:25.215 & $-$63:21:17.85 & 0.55$\pm$0.02 & -34.02$\pm$0.01 & 0.83$\pm$0.03 & 0.49$\pm$0.03 & 4.5-5.7 & 0.01-0.02   \\
Th12A$^c$	& 11:38:25.164 & $-$63:20:34.31 & 0.64$\pm$0.01 & -31.52$\pm$0.01 & 1.34$\pm$0.03 & 0.91$\pm$0.04 & 8.3-10.6 & 0.04-0.05  \\
Th37	& 11:38:24.163 & $-$63:20:11.29 & 0.89$\pm$0.02 & $-$31.466$\pm$0.005 & 0.49$\pm$0.01 & 0.46$\pm$0.02 & 4.2-5.4 & 0.02-0.03  \\
Th12B$^c$	& 11:38:24.132 & $-$63:20:31.04 & 0.67$\pm$0.05 & $-$32.49$\pm$0.02 & 0.64$\pm$0.06 & 0.46$\pm$0.07 & 4.2-5.4 & 0.01  \\
Th32	& 11:38:23.852 & $-$63:21:31.42 & 0.74$\pm$0.02 & $-$32.705$\pm$0.007 & 0.49$\pm$0.02 & 0.38$\pm$0.02 & 3.5-4.5 & 0.02  \\
Th34B$^c$	& 11:38:23.316 & $-$63:21:13.90 & 0.46$\pm$0.04 & $-$33.21$\pm$0.04 & 1.0$\pm$0.1 & 0.49$\pm$0.09 & 4.5-5.7 & 0.01  \\
Th36	& 11:38:22.901 & $-$63:20:21.68 & 0.86$\pm$0.02 & $-$31.609$\pm$0.006 & 0.75$\pm$0.02 & 0.69$\pm$0.02 & 6.3-8.1 & 0.04-0.05 \\
Th31	& 11:38:22.893 & $-$63:21:31.37 & 1.58$\pm$0.02 & $-$23.004$\pm$0.004 & 0.69$\pm$0.01 & 1.16$\pm$0.03 & 10.6-13.6 & 0.07-0.10  \\
Th30	& 11:38:22.774 & $-$63:21:36.32 & 0.79$\pm$0.02 & $-$23.55$\pm$0.01 & 0.87$\pm$0.03 & 0.73$\pm$0.04 & 6.7-8.6 & 0.02-0.03  \\ 
Th35	& 11:38:22.422 & $-$63:21:09.02 & 0.81$\pm$0.05 & $-$25.00$\pm$0.02	& 0.61$\pm$0.04 & 0.52$\pm$0.07 & 4.8-6.1 & 0.01  \\
Th13$^b$	& 11:38:21.83 & $-$63:20:54.45 & 0.46$\pm$0.01 & $-$24.65$\pm$0.01 & 1.04$\pm$0.03 &   &  &    \\
   &      &      & 0.10$\pm$0.01 & $-$23.56$\pm$0.04 & 0.57$\pm$0.09 & 0.29$\pm$0.01 & 2.6-3.4 & 0.04-0.05  \\
Th14	& 11:38:20.937 & $-$63:20:45.56 & 3.82$\pm$0.02 & $-$25.238$\pm$0.002 & 0.680$\pm$0.004 & 2.77$\pm$0.03 & 25.2-32.4 & 0.36-0.46  \\
Th15	& 11:38:20.820 & $-$63:20:55.35 & 1.28$\pm$0.03 & $-$22.734$\pm$0.008 & 0.64$\pm$0.02 & 0.86$\pm$0.04 & 7.9-10.1 & 0.06-0.07  \\ 
Th27	& 11:38:20.499 & $-$63:21:15.54 & 0.65$\pm$0.05 & $-$27.82$\pm$0.02 & 0.56$\pm$0.05 & 0.38$\pm$0.06 & 3.5-4.5 & 0.01  \\
Th26	& 11:38:20.485 & $-$63:21:18.39 & 0.70$\pm$0.06 & $-$32.95$\pm$0.02 & 0.39$\pm$0.04 & 0.29$\pm$0.05 & 2.6-3.4 & 0.004-0.005 \\
ALMA04	& 11:38:20.187 & $-$63:20:02.78 & 0.79$\pm$0.02 & $-$30.915$\pm$0.007 & 0.62$\pm$0.02 & 0.52$\pm$0.02 & 4.8-6.1 & 0.03-0.04  \\
ALMA08	& 11:38:18.909 & $-$63:20:03.67 & 1.32$\pm$0.02 & $-$30.287$\pm$0.004 & 0.533$\pm$0.009 & 0.75$\pm$0.02 & 6.8-8.8 & 0.06-0.07  \\
Th24	& 11:38:18.832 & $-$63:21:44.06 & 0.81$\pm$0.02 & $-$21.427$\pm$0.007 & 0.57$\pm$0.02 & 0.50$\pm$0.03 & 4.5-5.8 & 0.03  \\
Th23	& 11:38:18.076 & $-$63:21:39.97 & 0.97$\pm$0.06 & $-$22.707$\pm$0.016 & 0.53$\pm$0.04 & 0.56$\pm$0.07 & 5.0-6.4 & 0.01  \\
ALMA10	& 11:38:17.349 & $-$63:20:09.95 & 0.70$\pm$0.03 & $-$21.34$\pm$0.01 & 0.59$\pm$0.03 & 0.45$\pm$0.04 & 4.1-5.2 & 0.02-0.03  \\
%Th01	& 11:38:16.616 & $-$63:21:05.83 & 194$\pm$2 & $-$19.78$\pm$0.01 & 2.31$\pm$0.03 & 954$\pm$9 \\
Th01C$^a$	& 11:38:16.921 & $-$63:21:31.13 & 1.60$\pm$0.04 & $-$20.11$\pm$0.04 & 2.88$\pm$0.08 & 4.93$\pm$0.3 & 45.0-57.7 & 1.59-2.04  \\
Th01A$^a$	& 11:38:16.796	& $-$63:20:55.27 & 6.24$\pm$0.03 & $-$19.743$\pm$0.005 & 2.07$\pm$0.01 & 13.76$\pm$0.1 & 125.4-161.0 & 21.71-27.88  \\
Th01B$^a$	& 11:38:15.167 & $-$63:21:13.17 & 5.29$\pm$0.03 & $-$24.740$\pm$0.002 & 0.796$\pm$0.006 & 4.49$\pm$0.06 & 40.9-52.5 & 6.40-8.22  \\
Th22	& 11:38:14.595 & $-$63:21:34.91 & 0.6$\pm$0.02 & $-$25.14$\pm$0.01 & 0.67$\pm$0.02 & 0.43$\pm$0.03 & 3.9-5.0 & 0.02-0.03  \\
Th17	& 11:38:13.181 & $-$63:20:49.61 & 1.24$\pm$0.07 & $-$25.78$\pm$0.01 & 0.51$\pm$0.03 & 0.67$\pm$0.08 & 6.1-7.9 & 0.01  \\
Th16	& 11:38:12.947 & $-$63:20:37.15 & 0.69$\pm$0.03 & $-$28.66$\pm$0.01 & 0.51$\pm$0.03 & 0.38$\pm$0.04 & 3.4-4.4 & 0.02  \\
Th18	& 11:38:11.784 & $-$63:20:54.58 & 0.82$\pm$0.02 & $-$25.57$\pm$0.01 & 0.88$\pm$0.02 & 0.77$\pm$0.04 & 7.0-9.0 & 0.07-0.09  \\
Th20	& 11:38:11.406 & $-$63:21:21.97 & 1.20$\pm$0.06 & $-$23.93$\pm$0.01 & 0.59$\pm$0.03 & 0.82$\pm$0.07 & 7.4-9.6 & 0.01  \\
\hline
\multicolumn{9}{c}{Block B2} \\
\hline
ALMA02	& 11:38:22.326 & $-$63:21:25.57 & 0.76$\pm$0.05 & $-$9.49$\pm$0.02 & 0.77$\pm$0.06 & 0.63$\pm$0.08 & 5.7-7.4 & 0.01  \\
ALMA03	& 11:38:22.283 & $-$63:21:08.53 & 0.45$\pm$0.04& $-$7.39$\pm$0.05 & 1.1$\pm$0.1 & 0.55$\pm$0.1 & 5.0-6.4 &  0.01 \\
ALMA06	& 11:38:19.657 & $-$63:21:42.37 & 0.60$\pm$0.02 & $-$10.82$\pm$0.01 & 0.71$\pm$0.03 & 0.45$\pm$0.03 & 4.1-5.3 & 0.01-0.02  \\
\hline
\multicolumn{9}{c}{Block B3} \\
\hline
Th40	& 11:38:31.013& $-$63:20:22.19 & 0.47$\pm$0.04 & 45.11$\pm$0.02 & 0.41$\pm$0.04 & 0.20$\pm$0.04 & 1.8-2.3 & 0.004-0.005  \\
ALMA00	& 11:38:26.667 & $-$63:20:23.71 & 0.40$\pm$0.01 & 45.495$\pm$0.007 & 0.84$\pm$0.02 & 0.18$\pm$0.01 & 1.6-2.1 & 0.16-0.21  \\
ALMA01	& 11:38:25.570 & $-$63:19:57.62 & 0.54$\pm$0.08 & 45.81$\pm$0.02 & 0.30$\pm$0.05 & 0.17$\pm$0.05 & 1.6-2.0 & 0.002-0.003  \\
ALMA05	& 11:38:20.017 & $-$63:20:37.14 & 0.82$\pm$0.07 & 46.84$\pm$0.01 & 0.27$\pm$0.03 & 0.23$\pm$0.04 & 2.1-2.7 & 0.003-0.004  \\
ALMA07	& 11:38:19.479 & $-$63:20:30.80 & 0.28$\pm$0.01 & 48.022$\pm$0.009 & 0.62$\pm$0.02 & 0.19$\pm$0.01 & 1.72.2 & 0.03-0.04  \\
ALMA09	& 11:38:17.577 & $-$63:20:29.78 & 0.55$\pm$0.01 & 48.189$\pm$0.006 & 0.78$\pm$0.02 & 0.45$\pm$0.02 & 4.1-5.3 & 0.06-0.08  \\
ALMA11	& 11:38:14.750 & $-$63:20:05.91 & 0.10$\pm$0.01 & 48.98$\pm$0.01 & 1.17$\pm$0.03 & 0.13$\pm$0.01 & 1.2-1.5 & 0.13-0.16  \\
ALMA12	& 11:38:11.463 & $-$63:20:32.06 & 0.14$\pm$0.01 & 48.27$\pm$0.01 & 0.55$\pm$0.03 & 0.08$\pm$0.01 & 0.7-1.0 & 0.02-0.03  \\
\hline
\multicolumn{9}{p{0.8\linewidth}}{\raggedright Notes: \refcom{Line center velocities are expressed with respect to the LSR.} Last two columns have two values derived using excitation temperature of 11~K and 19~K, respectively.}\\
\multicolumn{9}{p{0.8\linewidth}}{\raggedright $^a$ Parts of the Th01, Th02, and Th03 globules. These appear as single entities \refcom{at} optical wavelengths \citep[although ][could distinguish at least two parts in Th01]{2003Reipurth} but show clear substructure in the ALMA molecular data. Parts are sorted by RA in this table.}\\
\multicolumn{9}{p{0.8\linewidth}}{\raggedright $^b$ Double-peaked spectra fit using two Gaussian. The reported \refcom{brightness is the integral (area under) of both kinematic components.}
}\\
\multicolumn{9}{p{0.8\linewidth}}{\raggedright $^c$ These are parts of individual entities/globules in the HST images (Th12 and Th34). Parts are sorted by RA in this table.
}
\end{tabular}
\end{table*}

\begin{figure*}
\centering
\includegraphics[width=1.15\linewidth]{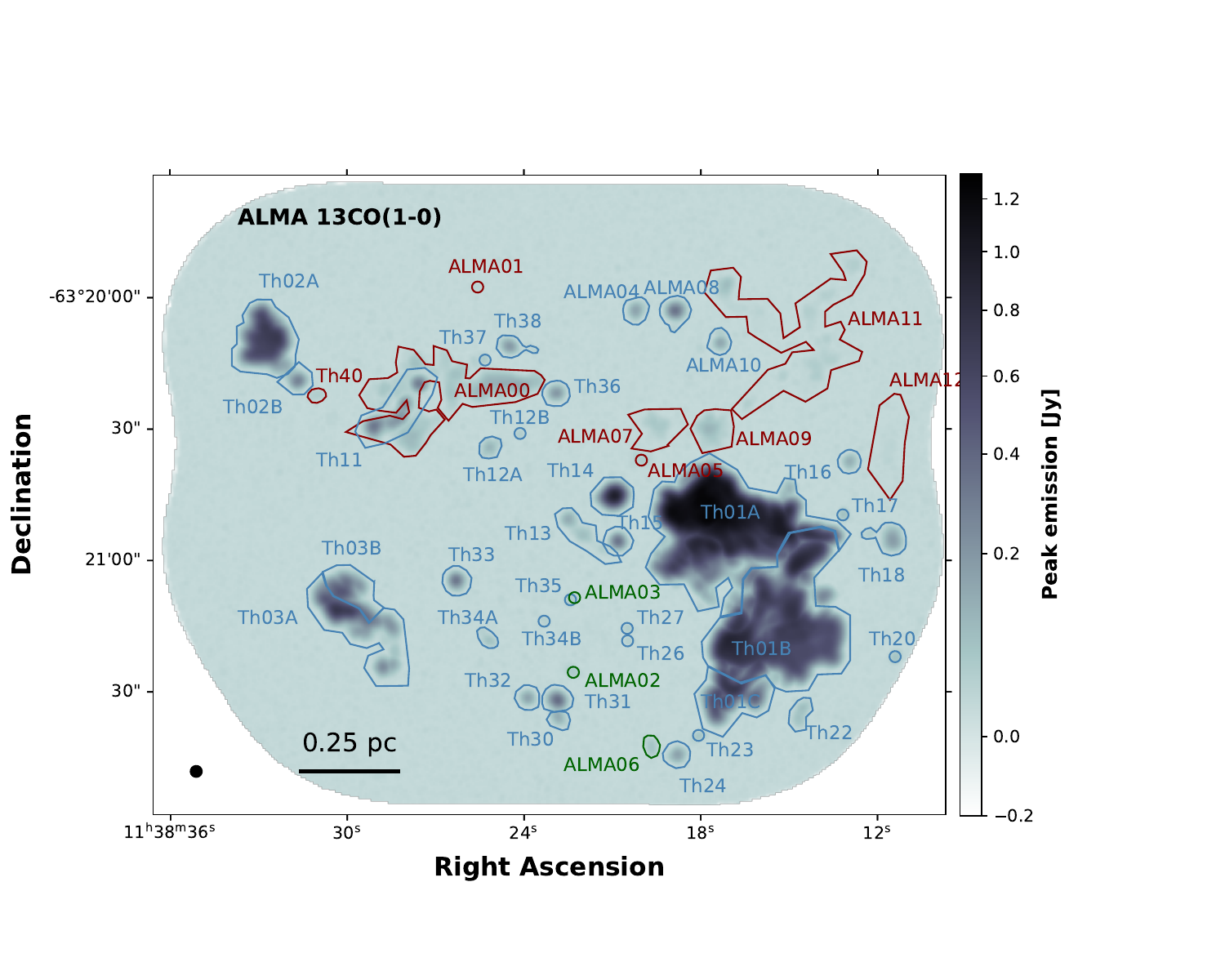}
\caption{$^{13}$CO~(1-0) peak emission (momentum 8) overlaid with symbols and polygons marking the detected molecular globules. Previously reported globules and splinters have \refcom{Th**} names, while newly detected ones follow the \refcom{ALMA**} convention. \refcom{Blue, green and red polygons are assigned to Blocks B1, B2, and B3, respectively.}}
\label{f:id}
\end{figure*}

%%%%%%%
\subsection{Continuum emission}
\label{sec:continuum}
\begin{figure}
\centering
\includegraphics[width=1.15\linewidth]{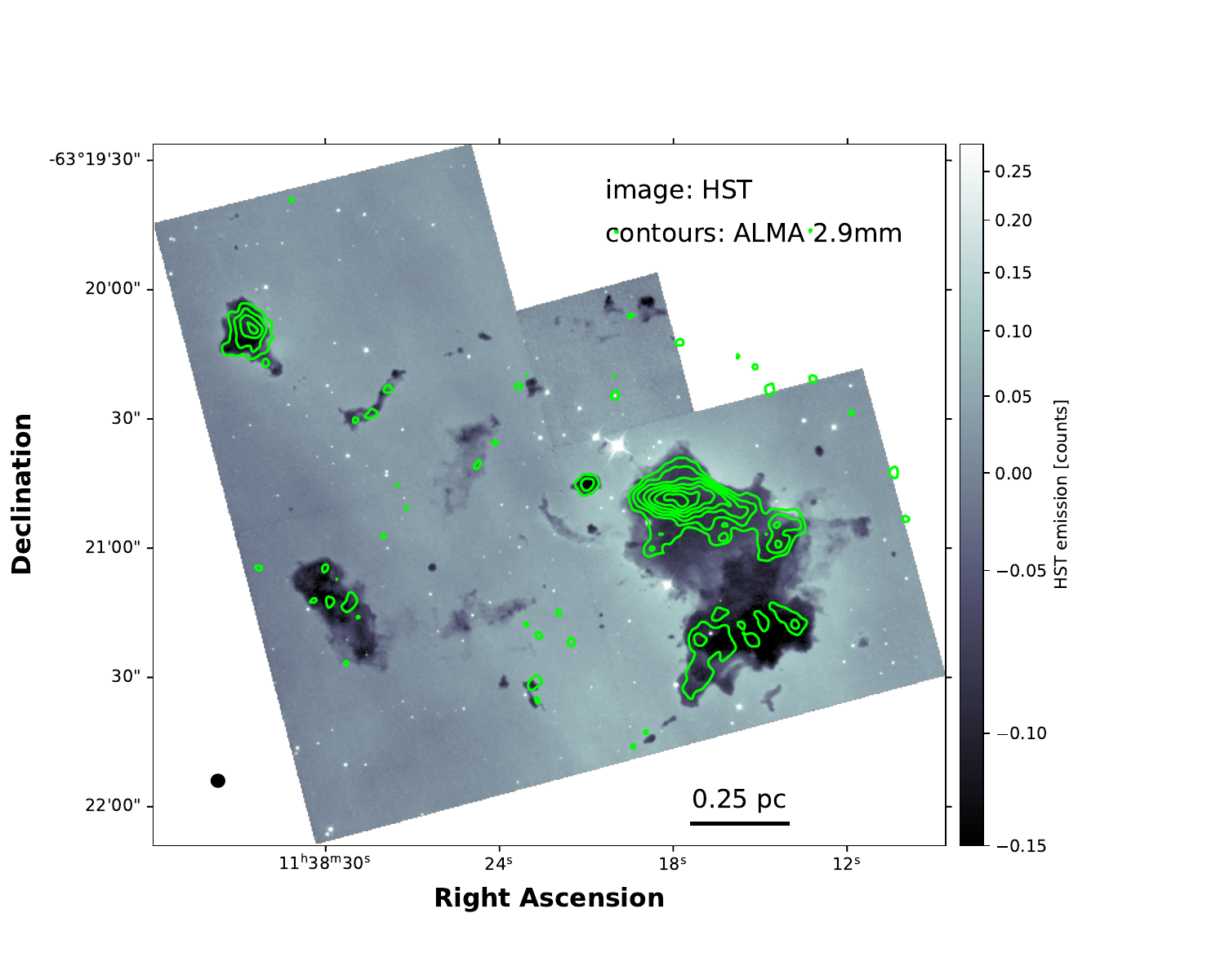}
\caption{HST image with 2.9~mm continuum contours overlaid at levels 3, 6, 9, 12, and 15$\sigma$, where $\sigma=21$\mujy is the rms noise level of the image.}
\label{f:cont}
\end{figure}

Figure \ref{f:cont} presents an image of the 2.9~mm continuum emission. There are four main sources associated with the globules included in Table \ref{t:continuum}. The strongest source dominating the emission spatially coincides with Th01A, peaking at  \dechms{11}{38}{17}{7338}, \decdms{63}{20}{48}{97} with an intensity of 485\mujy. 

We derived masses for these sources assuming an optically thin emission and using the expression $M=R\cdot D^2\cdot S_{\nu}\cdot[\exp(h\cdot\nu/(k\cdot T))-1]/(\nu^3\cdot\kappa)$. In the last equation, $R$ is the gas-to-dust ratio --here taken as 100--, $D$ the distance from Earth, $\nu$ the observed frequency --104.6~GHz--, $T$ is the dust temperature --that we consider ranging between 11~K and 19~K--, and the dust opacity $\kappa=\kappa_{\rm 1.3mm}\cdot(\nu/230.77)^{\beta}$. For the latter, we adopt $\kappa_{\rm 1.3mm}=0.65$ and $\beta=1.6$ for compact grains under typical ISM conditions \citep{1994Ossenkopf}. From this, we also estimate column and volume densities. The results (Table \ref{t:continuum}) roughly agree (mainly the lower bounds) with the values derived using the \tco emission. Th01A has a mass \bo{$\sim23-45$\msun}, while the compact continuum source harbored within Th03A is just about \bo{0.3-0.5\msun}.

\begin{table*}
\caption{Parameters derived from the dust emission.}
\label{t:continuum}
%\centering
\small
\begin{tabular}{@{}lccccccc@{}}
\hline\hline
\refcom{Object} & $S_{\nu}~^a$ & Size & \bo{n$_{\rm H_2}$} & \bo{Mass} & \bo{$\Lambda$} & \bo{$\nu$} & \bo{$t_{\rm ff}$} \\
      & [mJy] & [arcsec$^2$] & [$10^4\times $cm$^{-3}$] & [\msun] & & & [$10^5$~yr] \\
\hline \hline
Th01A	& 9.123 & 617.2 & 1.6-0.8 & 45-23 & 2.7-1.5 & 137.0-22.0 & 2.6-3.7 \\
Th01B-Th01C	& 1.980 & 264.3 & 1.2-0.7 & 9.7-5.1 & 1.5-0.9 & 25.8-4.6 & 3.0-4.1 \\
Th02A	& 1.421 & 108.1 & 3.4-1.8 & 6.9-3.6 & 1.7-0.9 & 30.9-5.2 & 1.8-2.5 \\
Th14	& 0.240 & 23.7 & 5.6-3.0 & 1.2-0.6  & 1.0-0.5 & 6.9-1.1 & 1.4-2.0 \\
Th03A$^b$& 0.105 & 15.0 & 4.9-2.6 & 0.5-0.3 & 0.7-0.4 & 2.7-0.5 & 1.5-2.0 \\
\hline
\multicolumn{8}{p{0.57\linewidth}}{\raggedright {\bf Note:} The two values for densities, masses, $\Lambda=L/\lambda_J$ and $\nu=M/M_J$ parameters (see Section \ref{sec:virial}), and free-fall times are derived using dust temperatures 11~K and 19~K, and appear in that order, respectively.} \\
\multicolumn{8}{p{0.57\linewidth}}{\raggedright $^a$ Fluxes are measured within a $2.5\sigma$ continuum emission level.}\\
\multicolumn{8}{p{0.57\linewidth}}{\raggedright $^b$ This is a \refcom{marginally resolved} and weak continuum source within the Th03 globule.}\\
\end{tabular}
\end{table*}

%%%%%
\subsection{Jeans' analysis of the globules}
\label{sec:virial}
Next we present an analysis of the stability of the globules against gravitational collapse by calculating their Jeans' length and mass using the classical expressions for spherical cores:
$\lambda_J=c_s/\sqrt{G\cdot\mu\cdot m_{\rm H}\cdot n}$
and $M_{\rm J}=(4\pi/3)\cdot \lambda_{\rm J}\cdot \mu\cdot m_{\rm H}\cdot n$. In these expressions $c_s=\sqrt{k\cdot T/(\mu\cdot m_{\rm H})}\approx0.18-0.24$\kms is the sound speed of the \tco at \bo{$T=$11-19~K}, with $\mu=2.8$ the mean molecular weight of the gas \citep[e.g., ][]{2012Crutcher}, and $m_{\rm H}=1.67\times10^{-24}$~g the atomic hydrogen mass. From the Jeans' length and mass we built the dimensionless parameters $\Lambda=L/\lambda_{\rm J}$ and $\nu=M/M_{\rm J}$ to compare the characteristic length and mass of the globules \bo{extracted from the \tco analysis}.  
Values of $\Lambda$ and/or $\nu$ larger than unity suggest a gravitational collapse. The median values for $\Lambda$ are \bo{$\approx$0.2}, and range between \bo{0.02 and 0.03 for $\nu$}, well below \refcom{1} in both cases. \refcom{This indicates that most of the globules are stable against gravitational collapse, in agreement with results from similar globules in the Rossette, NGC7822, IC1805 and Carina analyzed by \cite{2015Haworth}.}
However, for Th01A, Th01B-Th01C (which cannot be spatially separated in the continuum image, because they  overlap in projection along the line of sight), Th02A, Th03A, and Th14, at least one of the parameters is larger than \refcom{1} (Table \ref{t:continuum}), indicating that it is possible that some of these globules have the conditions to gravitationally collapse. 
For these cores, we repeated the Jeans' analysis focusing on the densest parts detected in continuum. We derive lengths $L$ assuming spherical sources whose size is defined by the 2.5$\sigma$ (52.5\mujy) contour of the dust emission (Figure \ref{f:cont}). Table \ref{t:continuum} \refcom{lists} the outcome of this analysis, suggesting that these sources, all detected in continuum, are prone to collapse (especially Th01A) in less than about 500,000~years.  

%%%
\subsection{Inner structure of Thackeray-1}
\label{sec:structure}
\begin{figure*}
\centering
\includegraphics[width=1.\linewidth]{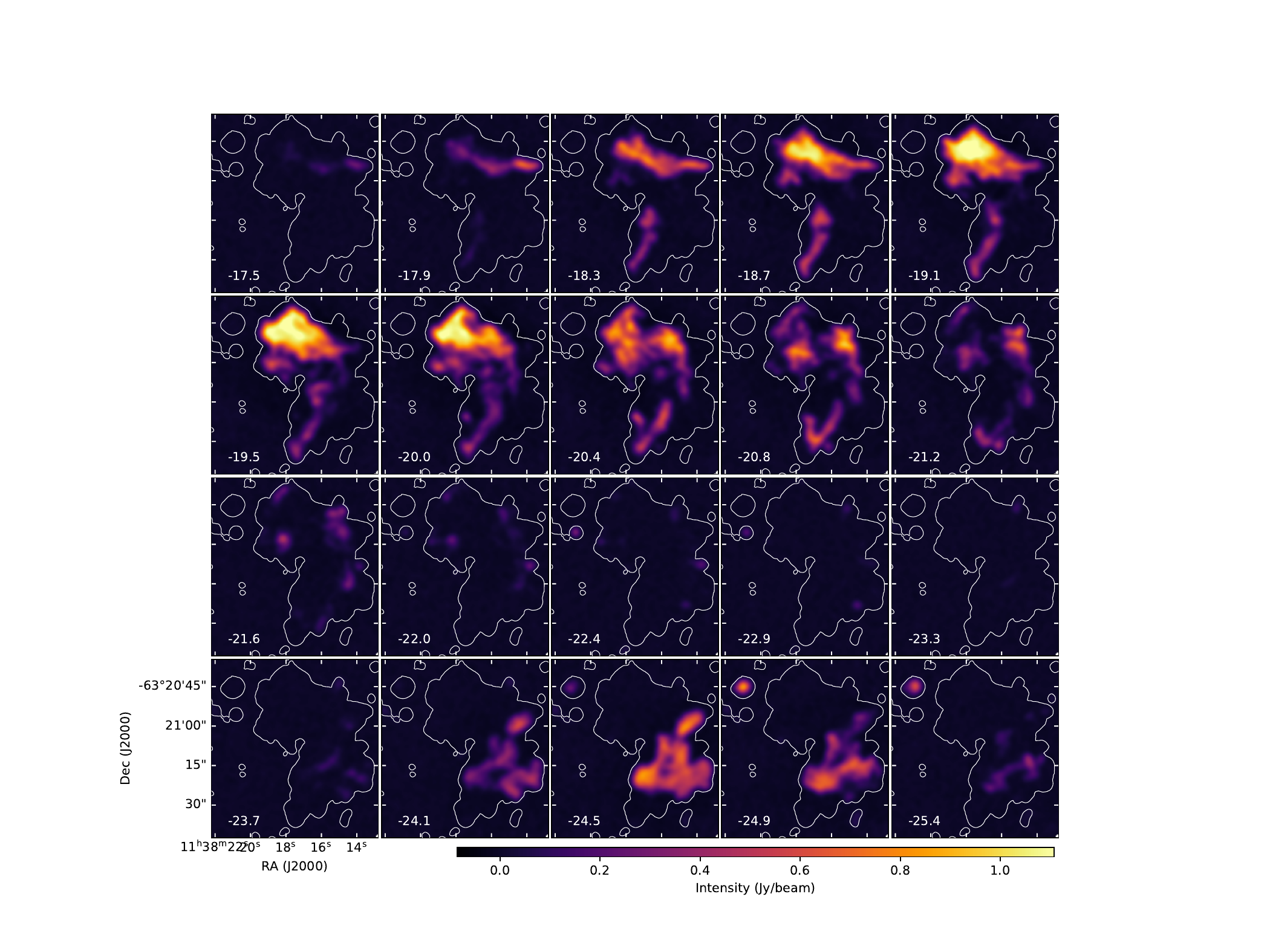}
\caption{$^{13}$CO\,(1-0) velocity cube of the Thackeray-1 globule with one contour overlapped from the $^{13}$CO moment 8 image at 0.03\jy. The line-of-sight velocity,  $v_{\rm lsr}$, in \kms appears at the bottom left in each channel.  The FoV also covers Th14 to the northeast.}
\label{f:cube01}
\end{figure*}

Figure \ref{f:cube01} presents the $^{13}$CO velocity cube towards Thackeray-1. The emission in the channel maps presents clumpy shoe-lace and blob morphologies. From $-$17.5\kms to $-$21.2\kms the structures comprising Th01A and Th01C, which seem to be connected, make an overall T-shape. The northern east-west branch of the T-shape, associated to Th01A, is brighter and harbors a continuum source. The southern north-south branch of the T-shape, mainly associated with Th01C, has a prominent filamentary shape roughly pointing to the position of HD~101205, reminiscent of the photoionized pillars found in M16 or Carina. This filament shows an east-west velocity gradient toward more blueshifted velocities. The T-shape structure \refcom{is} well separated kinematically from the more blueshifted emission stemming from Th01B (from $-$23.7\kms to $-$25.4\kms) and Th14 (from $-$24.5\kms to $-$25.4\kms). Th01B shows a more intricate network structure, with a southeast-northwest filament that extends to the west up to Th18 (not seen in the ALMA FoV). Th14 (northwest source) presents a clear southeast-northwest velocity gradient, which may be indicative of rotation, although with the data at hand, we cannot rule out other possible origin, such as rocket acceleration or Rayleigh-Jeans instability, for instance.

\subsection{Other molecular transitions}
\label{sec:other_lines}
\begin{figure}
\centering 
\begin{minipage}{0.55\textwidth}
\begin{subfigure}{\linewidth}
\includegraphics[width=\textwidth, trim= 0cm 1cm 0cm 1.5cm, clip]{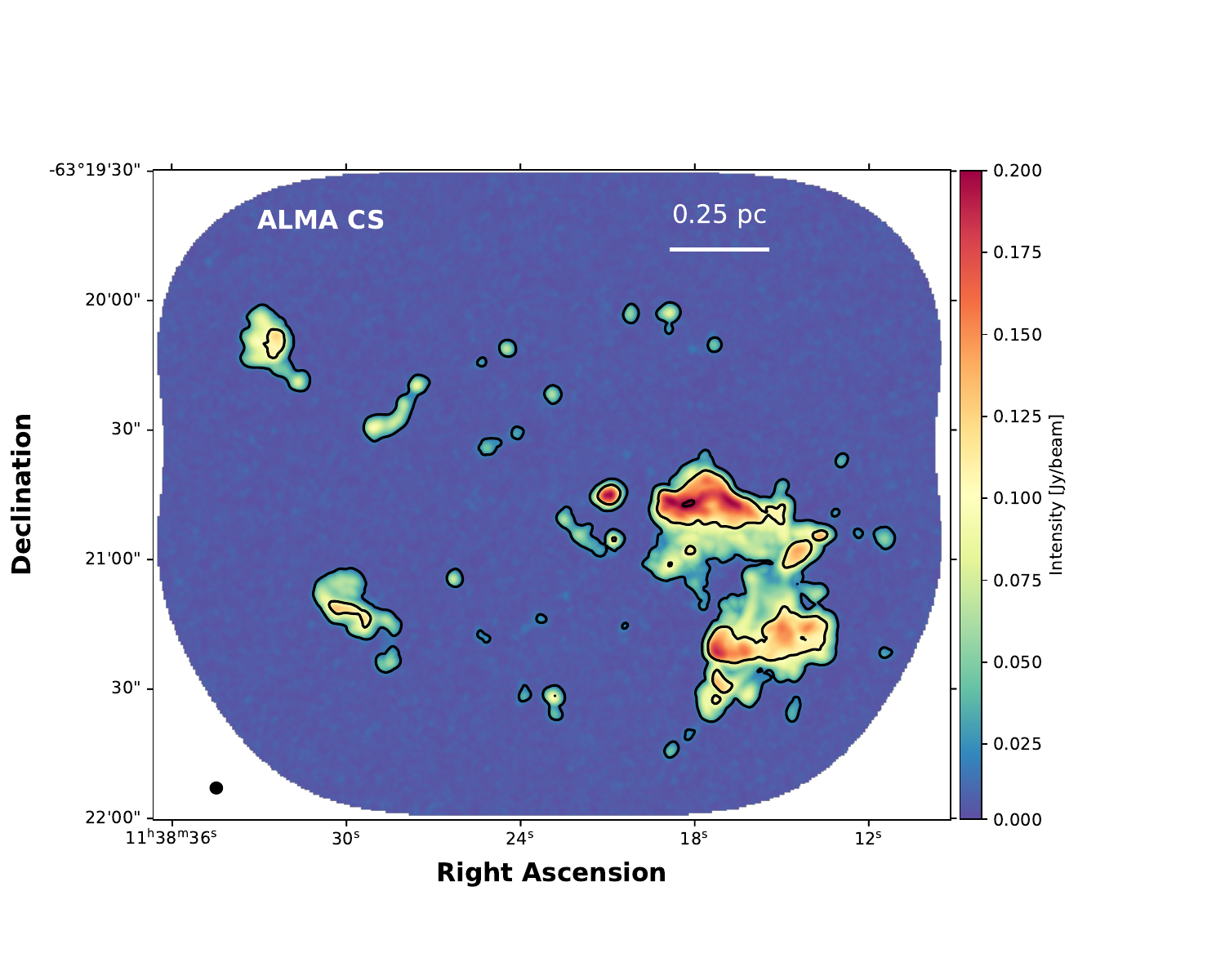}
\end{subfigure}
%\vspace*{-0.5cm}
\begin{subfigure}{\linewidth}
\includegraphics[width=\textwidth, trim= 0cm 1cm 0cm 1.5cm, clip]{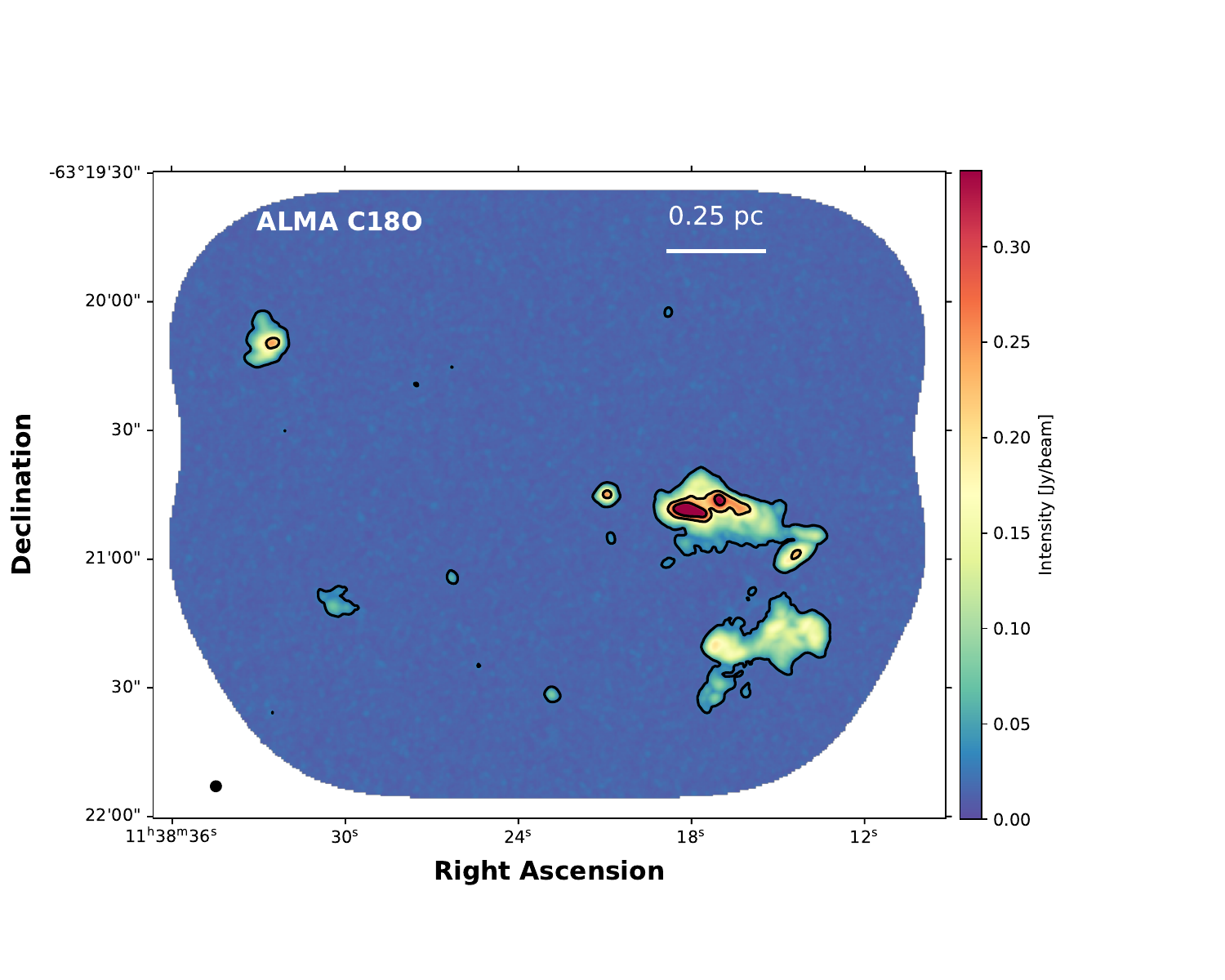}
\end{subfigure}
%\vspace*{-0.5cm}
\begin{subfigure}{\linewidth}
\includegraphics[width=\textwidth, trim= 0cm 1cm 0cm 1.5cm, clip]{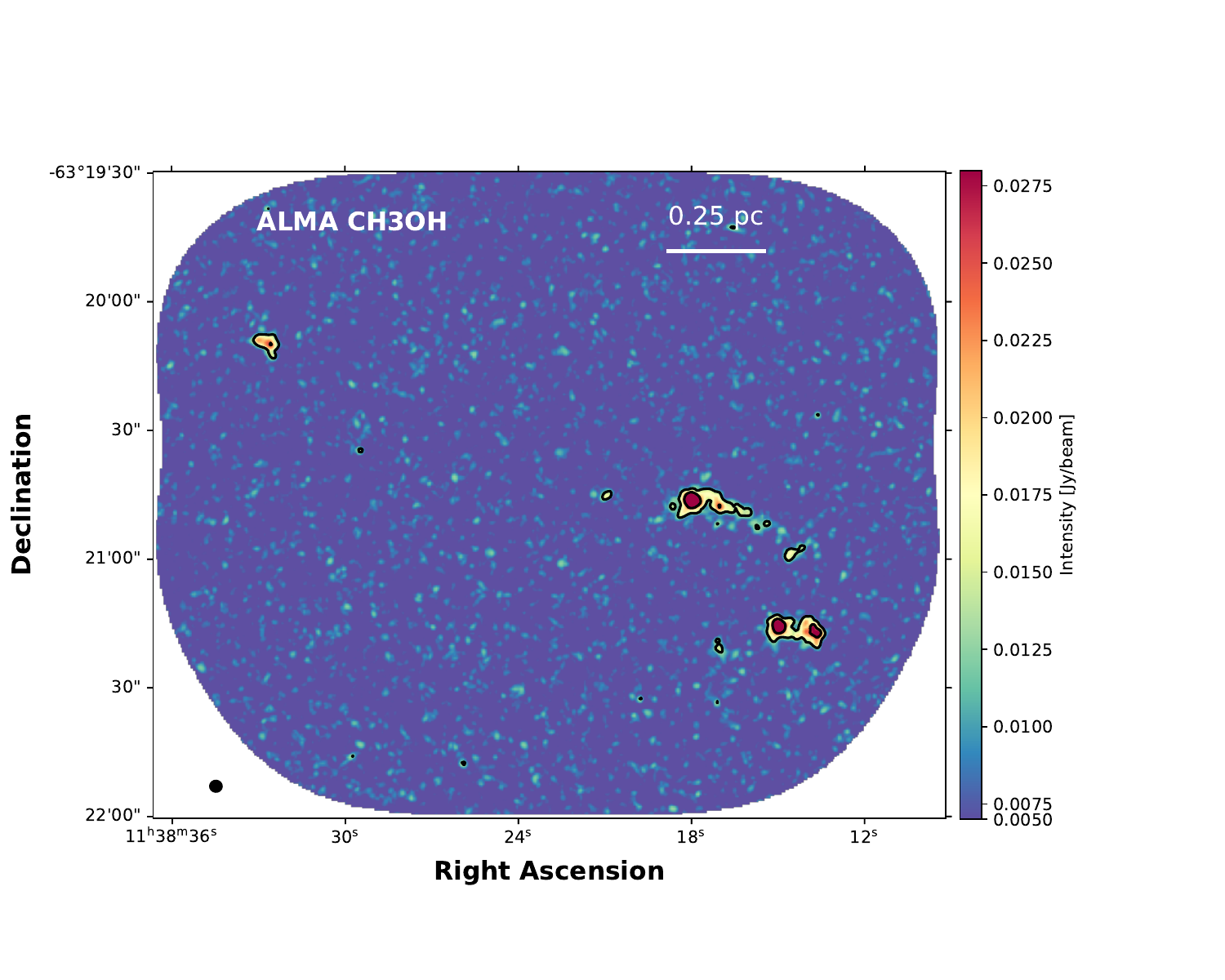}
\end{subfigure}
\end{minipage}
\caption{Moment 8 maps of the CS (top), C$^{18}$O (middle), and CH$_3$OH (bottom) transitions detected.}
\label{f:mom8s}
\end{figure}

In addition to the continuum and the \tco emission, the 2.9~mm ALMA observations presented in this work contain other lines (see Table \ref{t:lineid}). Figure \ref{f:mom8s} shows the moment 8 maps (derived by extracting the peak emission of the spectrum at every pixel) for the CS, C$^{18}$O, and CH$_3$OH lines. The CS emission is found in all the globules within the block B1, but it is not present in the intermediate velocity block B2 globules, nor in the redshifted block B3 globules. We may speculate that CS is destroyed in the illuminated surfaces of the globules. Since globules from blocks B2 and B3 are not detected in silhouette, it is plausible that we are \refcom{seeing} their illuminated faces directly, opposite to what happens with block B1 globules, \bo{or that they belong to background clouds that may not be associated with Collinder~249.}

C$^{18}$O and CH$_3$OH emission is more concentrated in the densest parts of the main globules (Th01, Th02, and Th03), including Th14 and Th31. \bo{In particular, the CH$_3$OH emission shows at least six unresolved compact sources associated with Th01A, Th01C, Th14, and Th02A. Two of them are associated with Th01A and two more with Th01C. All these sources may pinpoint the positions of dense cores/envelopes and coincide with the globules that have masses above the Jeans threshold.}

\begin{table}
\caption[]{Lines detected in the ALMA field.}
\label{t:lineid}
\centering
\begin{tabular}{lcccc}
\hline\hline
Transition & SPW & Rest Frequency & $E_{\rm up}$ & \bo{rms} \\
      &  & [GHz] & [K] & [\mjy] \\
\hline \hline
%U & spw0 & 110.21 \\
%SO(3,2-1,2) & spw1 & 99.299905 \\
C$^{18}$O~(1-0) & spw2 & 109.7821734 & 5.3 & 5.6 \\
\tco(1-0) & spw3 & 110.2013543 & 5.3 & 5.6 \\
%CH$_3$OD~(1,1,0-1,0,1)E & spw3 & 110.110.188860 \\
%HCCCN~(12-11) & spw3 & 110.1898 \\
CH$_3$OH~(2,0,2-1,0,1)A & spw4 & 96.741371 & 7.0 & 2.9 \\
CH$_3$OH~(2,0,2-1,0,1)E & spw4 & 96.744545 & 20.1 & 2.9 \\
CH$_3$OH~(2,-1,1-1,-1,0)E & spw4 & 96.755501 & 28.0 & 2.9 \\
$^{34}$SO~(3,2-2,1)E & spw5 & 97.715317 & 9.1 & 2.7 \\
%U & spw5 & 97.7393 \\
CS~(2-1) & spw6 & 97.9809533 & 7.1 & 2.7 \\
%SO$_2$~(33,3,31-32,4,28) & spw6 & 97.9940891 \\
%CH$_3$OCH$_3$~(16,3,14-15,4,11) & spw6 & 97.996174 \\
\hline
\end{tabular}
% \tablefoot{\\
% \tablefoottext{*}{Measured within a $2.5\sigma$ continuum emission level.}\\
% \tablefoottext{**}{This is a roughly compact and weak continuum source within the Th03 globule.
% }
% }
\end{table}

%\begin{figure*}
%\centering
%\begin{subfigure}[b]{0.3\textwidth} 
%\includegraphics[width=1.32\textwidth]{cs_mom8.pdf}
%\end{subfigure}
%\hfill 
%\begin{subfigure}[b]{0.3\textwidth}
%\includegraphics[width=1.32\textwidth]{c18o_mom8.pdf}
%\end{subfigure}
%\hfill
%\begin{subfigure}[b]{0.3\textwidth}
%\includegraphics[width=1.32\textwidth]{ch3oh_mom8.pdf}
%\end{subfigure}
%\caption{Moment 8 maps of the CS, C18O and CH3OH transitions detected.}
%\label{f:mom8s}
%\end{figure*}

\begin{figure}
\centering
\includegraphics[width=1.15\linewidth]{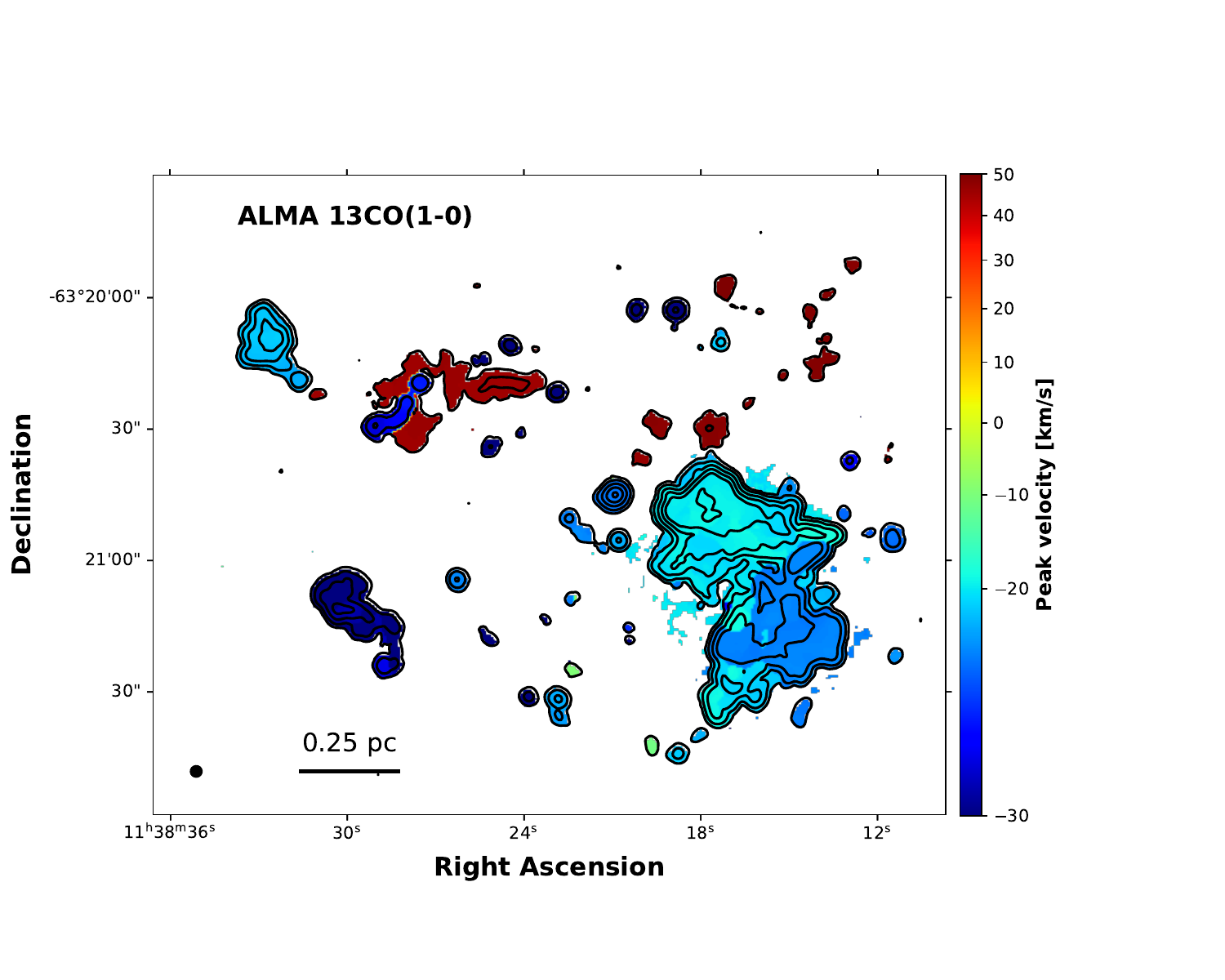}
\caption{$^{13}$CO (1-0) peak emission (moment~8) contours on top of the moment~9 image. \bo{While the moment~8 is built by finding the peak intensity of the spectrum associated with each spatial pixel of a velocity cube, the moment~9 map is built with the line-of-sight velocity, $v_{\rm lsr}$, corresponding to these intensity peaks.}}
\label{f:mom9}
\end{figure}

% \begin{figure}
% \centering
% \includegraphics[width=\linewidth]{Fig5.png}
% \caption{$^{13}$CO~(1-0) peak emission (momentum 8) contours on top of moment 2 image (weighted velocity dispersion).}
% \label{f:mom2}
% \end{figure}

%Inner kinematics and structure of Thackeray-1, Th-2, and Th-3

%%%%%%%

\section{Discussion}
\label{sec:discussion}
\subsection{Global kinematics}
\label{sec:globalkinematics}
\subsubsection{Line-of-sight velocity of the cloud}
\begin{figure*}
\centering
\includegraphics[width=1.0\linewidth]{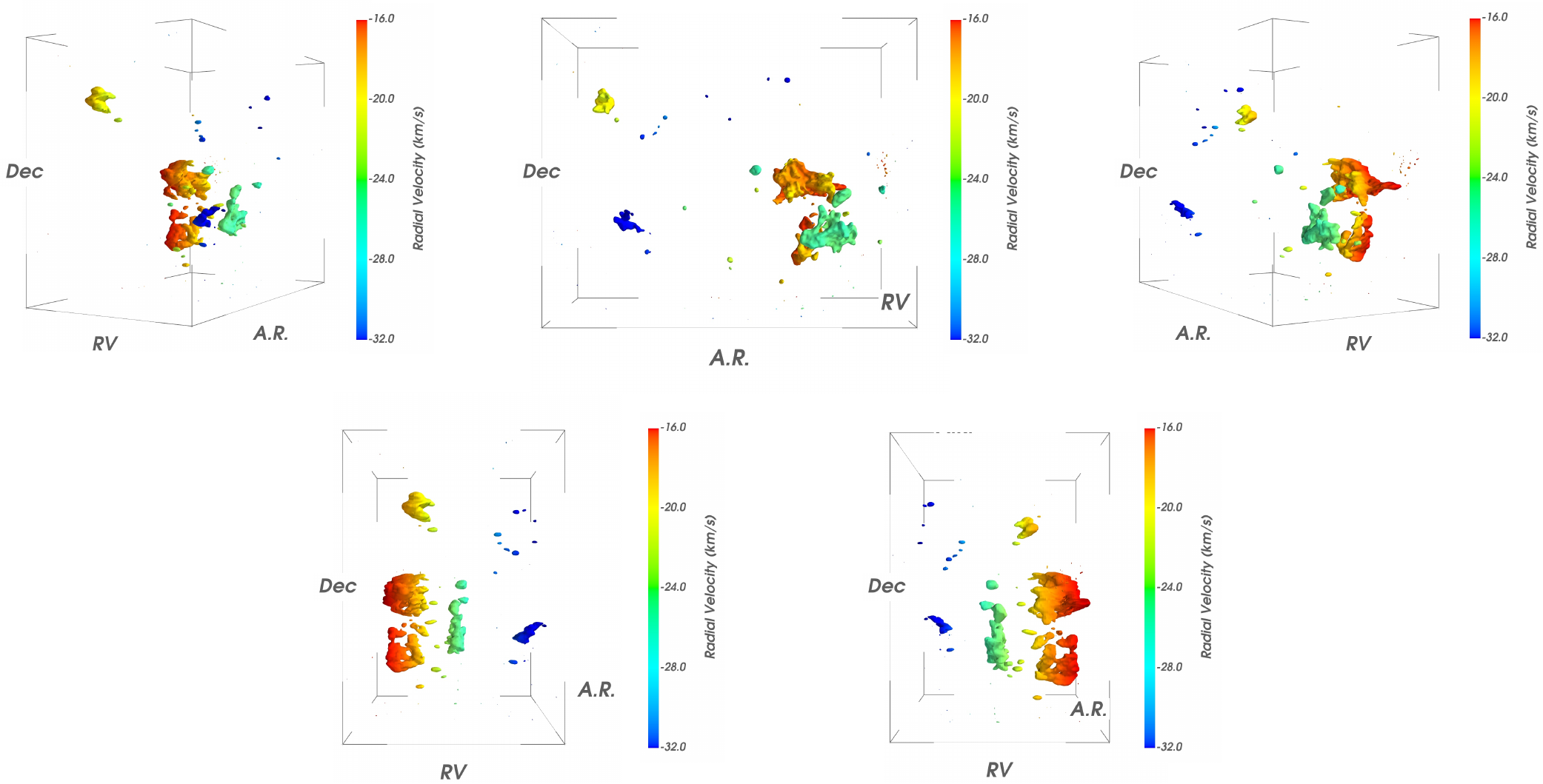}
\caption{3D Position-Position-Velocity (PPV) representations of the Thackeray's globules \tco emission. Each panel shows the view of the PPV space from a different perspective. Axes are RA, Dec and $v_{\rm lsr}$ (in a constrained range between $-$16\kms and $-$32\kms), and the color-scale also represents the line of sight velocity. The location of HD~101205 in the PPV space lies outside the boundaries of the box, as well as the globules of B2 and B3 blocks.}
\label{f:3D}
\end{figure*}

In the following, we discuss the adoption of a line-of-sight velocity ($v_{\rm lsr}$) for the original molecular cloud containing the Thackeray's globules. Collinder~249, the cluster of massive stars where HD~101205 belongs, have a heliocentric radial velocity of $-$14\kms, derived through [CaII] interstellar lines by \cite{2017Krelowski}, which transforms into a LSR velocity of $-$20.5\kms \citep[using the Sun's velocity relative to the LSR given by][]{2010Schonrich}. This $v_{\rm lsr}$ roughly agrees with the velocity extracted from the zero cut of the linear fits in the pv-plots presented in Figure \ref{f:pv}. However, this cut is coincident to the line-of-sight velocity of $-$16\kms of a low-density $^{12}$CO gas layer detected by \cite{1997Reipurth} that may comprise remnants of the gas of the original molecular cloud that hosted the cluster. With $-$16\kms as the cloud velocity, globules of the block B1 are blue-shifted from the ionizing stars, whereas globules of the blocks B2 and B3 are red-shifted. If all the globules are receding from the stellar cluster (pushed by winds, thermal pressure, and/or by the action of the rocket effect, see below), B1 globules should be in between the observer and the stars (at relative velocities $v_{\rm rel}= v_{\rm lsr}-v_{\rm cloud}<-4$\kms, according to the kinematics shown in Figure \ref{f:pv}), whereas B2 and B3 globules should be behind the stars (at relative velocities $>+9$\kms and $>+61$\kms, respectively). This scenario agrees with B1 globules seen in silhouette in the optical, while the rest of the globules are not detected by the HST and are only detected at millimeter wavelengths. Figure \ref{f:3D} shows a 3D representation of the location of the globules, assuming their position along the line of sight is proportional to their observed $v_{\rm lsr}$. 
%Finally, since block B2 globules are not detected in silhouette in optical, they should probably be in the background of the stars accelerating them at more redshifted velocities (pushed by winds, thermal pressure, and/or by the action of the rocket effect, see below). Thus, the velocity of the stellar cluster should be less than -11\kms. 
Taking into account these pieces of evidence, we henceforth adopt $-$16\kms as the cloud velocity and, in the following, give all line-of-sight velocities with respect to this value (i.e., we give $v_{\rm rel}$ velocities).

\subsubsection{Kinematics of the globules}
As seen in Figures \ref{f:pv} and \ref{f:mom9} the globules of the blocks B1 and B2 are distributed following linear velocity gradients. This would be unexpected if the globules were located on the surface of a spherical layer that expands at constant velocity. It would also be difficult to reconcile it with an origin of the globules from a large-scale turbulent medium bathed by ionizing radiation, with the expectation of more randomly distributed globule velocities \citep[e.g.,][]{2010Gritschneder,2013Tremblin}. In contrast, linear velocity gradients, as observed, can indicate that the globules are being accelerated to LSR velocities between $-$4\kms and $-$18\kms (B1 block), and $+$5\kms and $+$9\kms (B2 block). An alternative explanation for the velocity gradient is a large-scale Rayleigh-Taylor instability in a denser shell pushed by the expanding HII region, that will create pillar-like structures with the highest velocities at the their tips \citep{1997Reipurth}. Since we do not see clear velocity gradients along pillar-like structures (although there are some hints of this behavior in a few globules, such as parts of Th01A and Th01C), but isolated globules following a linear velocity gradient, we discuss a different working hypothesis.

In Section \ref{sec:acceleration} we will discuss the rocket effect mechanism as a means of accelerating the globules at a range of velocities. However, a complete scenario accounting for the kinematics of the region could in principle contemplate five stages. (i) A first stage that quickly establishes a Str\"omgren sphere, followed by (ii) the creation of a shell of compressed material driven out by thermal pressure within the HII region \citep[see e.g., ][and references therein]{1995Patel}. At the Str\"omgren radius, the velocity of this shell may be about $\sim10$\kms, approximately the sound speed in the ionized medium. (iii) The expanding shell eventually slows down as a result of the increase in the swept-up material from the surrounding gas. (iv) If the shell reaches the end of the original molecular cloud, the material accelerates again toward the less dense environment, and thermal pressure is no longer an effective driver for the material. (v) When the density of the shell, or any clumps that radiation may encounter within the HII region, is large enough, the material may be prone to undergo additional rocket-type acceleration. 
%This is specially effective after radiation escapes from the original cloud, and the drop in density exposes the clumps to the ionizing radiation from the O stars. 

Globules from the block B3 do not show a clear velocity gradient with projected distance, and their LSR velocity range, from \refcom{$+$61\kms to $+$65\kms}, is a factor of 3 larger than the largest receding velocity from the O stars of the rest of the globules. If B3 globules are impulsed by the radiation stemming from HD~101205 and HD~101191 (the brightest O stars in the cluster), they have undergone acceleration for a longer time, starting this process at a much closer range from the stars than the rest of the globules. An alternative is that block B3 globules do not relate to the expanding shell and acceleration process from these O stars, and they just serendipitously lie in projection with the other globules. Since the IC~2944 line-of-sight crosses the Carina spiral arm for kiloparsecs, there may inevitably be confusion with distant gas, and therefore, from now on we consider the B3 globules unrelated to Thackeray's globules and will not include them in further analysis.

\subsubsection{Distance between the globules and the ionizing stars}
\cite{1997Reipurth} estimated the projected separation for all Thackeray's globules. %Updated with the new derived distance the separation is 5.1~pc
\refcom{Using the new distance to the O-type stars (see Figure \ref{f:distances}), we find a separation of 5.1~pc} (note that the ALMA FoV does not include all the optical globules detected). If the group of globules were located at the surface of a sphere, the depth of the system would be similar to this separation, and then the block B1 globules would be at a distance of \bo{$\approx2.5$~pc} from the ionizing stars (half the total depth). This is almost four times larger than the Str\"omgren radius associated with HD~101205:
$R_s=\left[3S_*/(4\pi\alpha n^2)\right]^{1/3}\approx 0.64$~pc. In this expression $S_*$ is the ionization photon rate \citep[$\sim10^{49}$~photons~s$^{-1}$ for an O6V star,][]{2017Krelowski,1973Panagia}, $\alpha$ is the  recombination rate \citep[which we take as $3.1\times10^{-13}$~cm$^{-3}$~s$^{-1}$,][]{1978Spitzer}, and $n$ is the initial density (here assumed to be of order $\approx10^3$~cm$^{-3}$).

\refcom{However, the separation between the globules and the O stars is not clear. A different hypothesis may be assuming that the globule group has the same distance from the O stars as the large cometary cloud located to the northwest. This appears to be perpendicular to the line of sight and is separated $\sim13$~pc from the tip of the globule group (see Figure~\ref{f:distances}).} For the rest of this work, we make the assumption that the current distance to all the Thackeray's globules from the O stars is \refcom{between 2.5~pc and 13~pc, and use these as limiting values}.

% Discuss age of the cluster: 
% (1) Reipurth 2003: 1 Myr
% (2) Baume+14: 3Myr

\subsection{Acceleration of globules}
\label{sec:acceleration}

With the aforementioned assumptions we used the rocket equation to estimate the initial mass of each globule: $M/M_0=\exp^{-v/c_s}$, where $M$ and $M_0$ are the current and original mass of the globule respectively, $c_s$ is the speed of sound in the ionized medium \bo{(which we take as 10\kms)}, and $v$ is the current velocity of the globule \refcom{(approximated as $v_{\rm rel}$)}. A globule will reach a velocity about 7\kms\ after \refcom{losing} half of its original mass. 
Following the theoretical model of \cite{1998Mellema}, we can also estimate the rate of evaporated mass from the globules as a function of distance to the ionizing star: 
\refcom{
\begin{equation}\label{eq:massloss}
    \dot{M}=\frac{m S_*}{\left(1+\frac{\alpha_R S_* r_{\rm glob}}{12\pi c_s^2 R^2}\right)^{1/2}}\left(\frac{r_{\rm glob}}{2R}\right)^2\quad,
\end{equation} 
}
where $S_*$ is the ionizing photon rate of the stars (\refcom{taken as $10^{49}$~photons~s$^{-1}$}), $R$ and $r_{\rm glob}$ are the distance to the ionizing star and the size of the globule, \refcom{$m$ is the average mass per atom of the globule ($\approx1.33m_H$), and $\alpha_R$ is the case B hydrogen recombination rate coefficient ($\approx2.6\times10^{-13}$~cm$^3$~s$^{-1}$).} 

From this analysis, in the frame of the interstellar rocket, it is evident that the fastest globules today have been undergoing acceleration for a longer time or that their initial positions were closer to the ionizing stars.
\refcom{Using equation \ref{eq:massloss} from above, making $r_{\rm glob}\propto M^{1/3}$ for an uniform density globule \citep[note that there could be more sophisticated approaches such as ][based on numerical simulations]{2020Reiter}.}, and numerically integrating the rocket equation (in steps of \refcom{50} years) we obtained the velocity, mass, mass loss rate, velocity and distance to the stars as a function of time. Table \ref{t:acceleration} summarizes average values for different parameters per block globule. The focus of the Table is on three particular moments: the instant at which each globule started undergoing rocket acceleration ($t_{\rm start}$), the present time shown by the observations ($t=0$), and the moment at which \refcom{most of} the mass of each globule will be photoevaporated ($t_{\rm end}$). \refcom{If $R_{\rm end}$ reaches 50~pc the integration is also stopped and $t_{\rm end}$ is recorded at this earlier time instead.} The table contains the average values of the current size and mass, the ratio $M_0/M$ between the original and the current mass, the current value of the mass loss rate, the original mass of the globules \refcom{($M_0$)}, $t_{\rm start}$ (expressed as a negative value, since we consider zero time the present time), the distance at which the globules started the rocket acceleration process \refcom{($R_{\rm start}$)}, $t_{\rm end}$ (i.e., the remaining time until the globules are completely consumed by the ionizing radiation), the distance at which the evaporation will be completed \refcom{($R_{\rm end}$)}, and the velocity at this point \refcom{($v_{\rm end}$)}. \refcom{Some of the parameters have two entries in the table, corresponding to the different distances between the O stars and the globules contemplated (the limiting cases 2.5~pc and 13~pc, respectively). In the following, we refer to the values derived using 2.5~pc for a qualitative discussion of the results.}

\begin{table*}
\caption{Results from model of rocket--accelerated globules.}
\label{t:acceleration}
\small
%\centering
%\setlength{\tabcolsep}{2pt}
\begin{tabular}{@{}lcccccccccc@{}}
\hline\hline
Globule block & \refcom{$L$} & $M$ & $M_0/M$ & $\dot{M}_{\rm current}$ & $M_{0}$ & t$_{\rm start}$ & $R_{\rm start}$ & t$_{\rm end}$ & R$_{\rm end}$ & $v_{\rm end}$\\
      & [pc] & [\msun] &  & [$10^{-6}$~\msun~yr$^{-1}$] & [\msun] & [$10^{5}$~yr] & [pc] & [$10^{5}$~yr] & [pc] & [\kms] \\
\hline \hline
\multicolumn{11}{c}{Block Averages} \\
\hline
B1--massive$^a$ & 0.15 & 5.4  & 2.5 & 5.3/1.0 & 9.7 & $-$3.0/$-$15.7 & 1.2/6.2 & 28.9/34.3 & 50/50 & 23/15 \\
B1--light$^a$	& 0.04 & 0.03 & 3.5 & 0.8/0.4 & 0.11 & $-$0.5/$-$1.1 & 2.2/12.3 & 0.9/1.4 & 4.9/16.8 & 43/43 \\
B2	            & 0.03 & 0.01 & 2.1 & 0.4/0.06 & 0.02 & $-$0.2/$-$1.4 & 2.4/12.5 & 0.4/3.4 & 3.2/19.1 & 30/30 \\
%B3	        & 0.06 & 666.9 & 1.9 & 26.3 & $-$10.9 & 0.2 & 0.2 & 2.7 & 82  \\
\hline
\multicolumn{11}{c}{Individual Globules} \\
\hline
Th01A & 0.26 & 24.8  & 1.4 & 11.1/2.1 & 35.9 & $-$5.9/$-$31.0 & 1.3/6.6 & 59.7/66.8 & 50.0/50.0 & 10/6 \\
Th01B & 0.25 & 7.3   & 2.4 & 10.3/1.9 & 17.4 & $-$3.5/$-$18.3 & 0.7/3.6 & 30.2/32.0 & 50.0/50.0 & 18/13 \\
Th01C & 0.12 & 1.8   & 1.5 & 3.3/0.6 & 2.7 & $-$2.1/$-$11.6 & 2.0/10.6 & 30.1/41.2 & 50.0/50.0 & 20/12 \\
Th02A & 0.13 & 2.1   & 1.8 & 3.8/0.7 & 3.7 & $-$2.8/$-$14.7 & 1.7/8.5 & 29.2/36.9 & 50.0/50.0 & 20/12 \\
Th03A & 0.14 & 1.1   & 4.0 & 4.6/0.8 & 4.4 & $-$2.2/$-$11.6 & 0.7/3.7 & 16.9/19.3 & 50.0/50.0 & 33/22 \\
Th14  & 0.07 & 0.4   & 2.5 & 1.7/0.3 & 1.0 & $-$2.0/$-$10.9 & 1.5/7.8 & 18.3/24.3 & 50.0/50.0 & 32/18 \\
\hline
\multicolumn{11}{p{0.7\linewidth}}{\raggedright {\bf Note:} 
\refcom{The values in this table are averaged quantities per block of globules (top three rows) and values derived for some of the most massive globules (bottom rows). Columns showing two-value entries give the results corresponding to two different distances between the O stars and the globules: 2.5~pc and 13~pc, respectively.} 
Columns: (1) Label of block. \refcom{(2) Average of the characteristic diameter of  the globules in each block.} (3) Average of the current mass of the globules derived from the \tco observations. (4) Ratio between the original ($M_0$ at time $t_{\rm start}$) and the current mass ($M$ at time $t=0$) of the globules until the current observed speed is reached (average per block). (5) Mass loss rate at the present time $t=0$. (6) Original mass of the globule (i.e., mass at $t_{\rm start}$). (7) Time at which the globule started acceleration by the rocket effect ($t_{\rm start}$). (8) Distance to the accelerated globules from the O stars at $t=t_{\rm start}$. (9) Remaining time until the globule is completely photoevaporated ($t_{\rm end}$). (10) Distance from the protostar at time $t_{\rm end}$. (11) Velocity at time $t_{\rm end}$.}\\
\multicolumn{11}{p{0.75\linewidth}}{\raggedright $^a$ Block B1 statistics are split into two groups: \textit{massive} and \textit{light}. The group \textit{Massive} comprises globules Th01, Th02, Th03, and Th14, and the group \textit{light} comprises the rest of the B1 globules.}\\
\end{tabular}
\end{table*}

\refcom{These estimates} show certain interesting aspects:
\begin{itemize}
\item Block B1 globules have two separate regimes. The globules in possible gravitational collapse (ie., Th01, Th02, Th03, and Th14, or B1-massive in Table \ref{t:acceleration}) started to be accelerated approximately \refcom{300,000~years} ago, at a distance of \refcom{1.2~pc} away from the stars, on average. Before undergoing strong photo-illumination, these were cores of about \refcom{10\msun} on average (spanning a range \refcom{1-36\msun}) and currently have lost \refcom{40\% of} their original mass. Non-collapsing B1 globules (almost 70\% of all detected globules; designated as B1-light in Table \ref{t:acceleration}) started acceleration about \refcom{50,000~years} ago, at \refcom{2.2~pc} from the stars, and their original masses were about 0.1\msun. This could indicate that lighter globules and splinters were originally further away from the ionizing stars, or that they fragment and detach from the larger globules about \refcom{50,000~years} ago.

\item The three B2 globules had small masses \refcom{($M_0\sim 0.02$\msun)}, and currently display small velocities with respect to that of the ionizing stars. Their rocket acceleration started \refcom{20,000~yr ago, when they were 2.4~pc} away from the stars.
%\item B3 globules currently display very large velocities ($\sim 60$\kms on average). If they were accelerated by the interstellar rocket effect \bo{from the Collinder~249 cluster, they were originally much more massive ($M_0=26.3$\msun vs their current $M=0.06$\msun, on average)}. In this case, these globules were accelerated for \bo{109,000~yr}, starting the process very close to the stars, at \bo{0.2~pc}.
\end{itemize}

\refcom{Finally, let us make the caveat again that considering a different current separation between globules and O stars (i.e., 13~pc instead of 2.5~pc), these numbers would be scaled accordingly implying larger $t_{\rm start}$, and $R_{\rm start}$ values (see Table \ref{t:acceleration}).}

\subsection{Evaporation: globule survival}
With the same assumptions, we have used the numerical integration of the rocket equation to derive \refcom{the remaining lifetime.}
%how long it takes until the globules are photoevaporated. 
Table \ref{t:acceleration} provides an estimate of the remaining time ($t_{\rm end}$).  Provided that the \refcom{the current distance to the O stars is 2.5~pc}, most of the globules will last, on average, less than \refcom{1~Myr}. However, for the gravitational collapsing globules that still possess large mass reservoirs, the evaporation times are larger than \refcom{1.7~Myr}. With this model, the most massive Th01A, currently harboring about 25\msun, will last for \refcom{more than 6~Myr, going over $50$~pc away from the cluster of O stars, and reaching velocities of $\sim10$\kms.} This survival time is \refcom{ more than 1 order of magnitude} larger than its free-fall time (Table \ref{t:continuum}), which may suggest that it has enough time to condense new protostars. \refcom{The collapse of globules under strong ionization has also been shown feasible in numerical simulations \citep[e.g., ][]{2007Esquivel}.} 
It is worth noting that these protostars would inherit \refcom{the high velocity of the globule} and may become low-mass walkaways/runaways from Collinder~249. In this hypothesis, the interstellar rocket effect may be a plausible mechanism to accelerate stars to high-speeds that can eventually \refcom{escape} from their original birth places. However, as we shall see in the next section, the globule survival is threatened by other processes which speed up destruction, indicating that the runaway hypothesis is a difficult outcome.

\subsection{Fragmentation of the globules}
From the analysis of the detailed HST images of the main globules, \cite{2003Reipurth} identified a large population of splinters surrounding Th01. Since \refcom{photoevaporation}
%gas density evaporation
is proportional to the surface area, i.e., to $r^2$, while the mass is proportional to the volume, i.e., to $r^3$, it follows that the smallest splinters are destroyed in very short times. The fact that many are observed to exist suggests that many of these small bodies are detached from the larger ones through some fragmentation mechanism. \cite{2003Reipurth} concluded that if small globules are
continuously formed through fragmentation and they are destroyed through evaporation, then the relative efficiency of these processes must lead to the observed distribution of fragments.

As pointed out in the previous section, our analysis places most of the smallest globules \refcom{relatively} close to their current position only a few \refcom{tens of thousands} years ago, when they started to be propelled by the rocket effect. In contrast, the main globules have been accelerated from further away during much longer periods. This indicates a good agreement with a recent fragmentation and detachment from larger globules, and explains why, despite their short lifetimes, small globules and splinters are still observed.
\refcom{There is an exception to this conclusion: block B2 globules are not associated with any large globule, so in this case, fragmentation and detachment is not a good explanation. These globules, undetected at optical wavelengths, show lower velocities and are far less numerous (about 8\% than B1 globules). Under the rocket effect hypothesis, these are fast transient overdensities. Compared with the age of the ionizing stars in IC~2944 \citep[3~Myr, according to][]{2014Baume}, it is clear that their short lifetimes indicate: (i) that there is a large number of them to be destroyed, (ii) that they belonged to much larger bodies or clouds, or (iii) that they started being ionized much further away. Considering the largest distance from the O stars of $\sim13$~pc, their survival lifetime would be about 340,000~yr.}

\section{Summary}
\label{sec:summary}
We carried out an analysis of ALMA Band 3 data toward the Thackeray globule complex associated with the Collinder~249 cluster. We could identify most of the optical globules in the \tco images and study their individual kinematics in some detail. \refcom{We also discover 13 new globules undetected at other wavelengths. From the analysis of the molecular gas we also derived densities and masses. Most of the mass is concentrated in the few main extended globules, clearly detected in the 2.9~mm continuum emission image, and showing signs of gravitational collapse, as well. In particular we analyzed the inner structure of Thackeray-1, which shows two kinematically separate structures, Th01B and Th01A-Th01C, with the latter forming a T-shape with a north-south pillar-like feature pointing toward HD~101205, the dominant ionizing source in the region. We reported the detection of some other molecular emission lines including high density tracers such as \ceto and CH$_3$OH toward the most dense parts of the globules, which may again suggest the existence of soon-to-be gravitationally collapsing cores. We divided the globules in three different blocks according to their line-of-sight velocity.} The B1 block is blue-shifted ($v_{\rm rel}$ from \refcom{$-$4\kms to $-$18\kms}) with respect to the \refcom{cloud} velocity while the B2 and B3 blocks are red-shifted, with average velocities of +7\kms and +63\kms, respectively. Globules from blocks B1 and B2 seem to follow a linear velocity gradient, indicative of a very dynamical process. 

In the second part of the paper, we discussed the possibility that the velocity gradient along the globules may be explained due to the interstellar rocket acceleration. In this scenario B1 globules would lie between the observer and the ionizing stars, while B2 and B3 globules lie behind them. After \refcom{assuming two limiting cases for the current separation between the globules and the O stars (2.5~pc and 13~pc),} we used analytical expressions to make rough estimates of the Thackeray's globules acceleration.
\refcom{We find different outcomes for every globule. Depending on their current mass and velocity, we estimate a range of initial masses, times they have been accelerated, and initial distances from the O-type stars for the B1 and B2 blocks.} We also estimate the time it will take to evaporate the current globules.
Although most of the globules have remaining lifetimes less than \refcom{100,000~years}, the most massive globule, Th01A, could resist destruction for more than \refcom{3~Myr}. The short lifetimes and small sizes and masses of most of the globules and splinters in the region can be explained by recent fragmentation and detachment from the largest globules. The fragmentation process greatly speeds up photo-evaporation of the globules and opposes the collapse and formation of protostars within them. In the rare and extreme case that it happens, the newly formed stars may be accelerated at high speeds (up to \refcom{$\sim30$\kms}) as the globule is ejected away from its original location, potentially creating walkaway/runaway low-mass stars.

\bo{More work is required to confirm the role of the interstellar rocket effect in the kinematics of these types of globules. For starters, there are more globules in the southwest of \refcom{IC~2944}, located outside the ALMA and HST fields of view (Figure \ref{f:distances} in Section \ref{sec:distance}). It would be important to confirm that their kinematics and fragmentation history agree with the rocket acceleration scenario hypothesized in this work. Another experiment is to measure the kinematics of the ionized gas streaming off the globule through recombination line observations to test whether this may propel the globules. To test the hypothetical and extreme case of runaway formation, probing the dust emission from millimeter to near-infrared wavelengths with higher angular resolution and searching for protostellar seeds can reveal possible star-formation content within the globules (from pre-stellar to Class 0 protostars). Refinement of the toy model for photoevaporation presented here may include \refcom{a more accurate approach for the shrinking} of globules with time, fragmentation and detachment. All these effects may contribute to the evaporation time. Finally, globules akin to Thackeray's have been observed in prominent and classical HII regions such as M16, the Rosette nebula or the Carina nebula \citep[e.g.,][]{1996Hester,2003Smith,2007Gahm,2014Grenman}. Some of them already have evidence of high-velocity cloudlets that are worthy of closer inspection \citep[e.g.,][]{1986Meaburn}. Interferometric (sub)millimeter observations can pinpoint the kinematics of these globules to test whether our findings can be generalized to other classical star-forming regions.}

\section*{Acknowledgements}
\refcom{The authors want to thank the referee, Thomas J. Haworth, for comments and suggestions that helped to improve the manuscript.

M.F.L. acknowledges the warmth and hospitality of his colleagues at Instituto de Radioastronom\'ia y Astrofísica, UNAM, in Morelia. L.A.Z. and M.F.L. acknowledge financial support from CONACyT-280775, UNAM-PAPIIT IN110618, and IN112323 grants, Mexico.}

This paper makes use of the following ALMA data: ADS/JAO.ALMA\#2015.1.00908.S. ALMA is a partnership of ESO (representing its member states), NSF (USA) and NINS (Japan), together with NRC (Canada), MOST and ASIAA (Taiwan), and KASI (Republic of Korea), in cooperation with the Republic of Chile. The Joint ALMA Observatory is operated by ESO, AUI/NRAO and NAOJ.

This work also makes use of observations made with the NASA/ESA Hubble Space Telescope, obtained from the Data Archive at the Space Telescope Science Institute, which is operated by the Association of Universities for Research in Astronomy, Inc., under NASA contract NAS5-26555.
These observations are associated with program \#7381.

Finally, this work makes use of data from the European Space Agency (ESA) mission Gaia (https://www.cosmos.esa.int/gaia), processed by the Gaia Data Processing and Analysis Consortium (DPAC, https://www.cosmos.esa.int/web/gaia/dpac/consortium).
%%%%%%%%%%%%%%%%%%%%%%%%%%%%%%%%%%%%%%%%%%%%%%%%%%
\section*{Data Availability}
The calibrated data underlying this article is available from the corresponding author upon reasonable request.



\bibliographystyle{mnras}
\bibliography{biblio} % if your bibtex file is called example.bib

%\begin{thebibliography}{99}
%\end{thebibliography}

%%%%%%%%%%%%%%%%%%%%%%%%%%%%%%%%%%%%%%%%%%%%%%%%%%

%%%%%%%%%%%%%%%%% APPENDICES %%%%%%%%%%%%%%%%%%%%%

%%%%%%%%%%%%%%%%%%%%%%%%%%%%%%%%%%%%%%%%%%%%%%%%%%

% Don't change these lines
\bsp	% typesetting comment
\label{lastpage}
\end{document}